\documentclass[traditabstract,10pt]{aa}

\usepackage[T1]{fontenc}
\usepackage[english]{babel}
\usepackage{amsmath, amssymb, amsfonts, latexsym}
\usepackage[breaklinks, colorlinks, citecolor=blue]{hyperref} 
\usepackage{ifthen}
\usepackage[usenames,dvipsnames,table]{xcolor}
\usepackage{placeins}
\usepackage{txfonts}

\usepackage{graphicx}
\graphicspath{{Figures/}}
\usepackage{float}
\usepackage{subfigure}
\usepackage{epstopdf}
\usepackage{dblfloatfix}
\usepackage{morefloats}
\usepackage[skip=5pt]{caption}
 
\usepackage[nonamebreak]{natbib}
\bibpunct{(}{)}{;}{a}{}{,}
\providecommand{\sorthelp}[1]{}

\def\farcm{\ifmmode {^{\scriptstyle\prime}} \else $^{\scriptstyle\prime}$\fi}
\newcommand{\draft}{false}
\newcommand{\LCDM}{{$\rm{\Lambda CDM}$}}

\def\GHz{\ifmmode $\,GHz$\else \,GHz\fi}
\def\planck{\textit{Planck}}

\newcommand{\alm}{\ifmmode {\vec{a}_{\ell m}} \else $\vec{a}_{\ell m}$\fi}

\newcommand{\Mpc}{\mathrm{Mpc}}
\def\pdeg{\ifmmode $\setbox0=\hbox{$^{\circ}$}\rlap{\hskip.11\wd0 .}$^{\circ}
          \else \setbox0=\hbox{$^{\circ}$}\rlap{\hskip.11\wd0 .}$^{\circ}$\fi}

\newcommand{\CL}[1]{#1\,\%\ CL}

\newcommand{\lollipop}{{\tt LoLLiPoP}}
\newcommand{\hillipop}{{\tt HiLLiPoP}}
\newcommand{\lowl}{\mbox{low-$\ell$}}
\newcommand{\highl}{\mbox{high-$\ell$}}
\newcommand{\commander}{{\tt Commander}}
\newcommand{\hlp}{{hlp}}
\newcommand{\plik}{{plik}}
\newcommand{\lowT}{{lowT}}
\newcommand{\lowE}{{lowE}}
\newcommand{\lolE}{{lowlE}}
\newcommand{\lolB}{{lowlB}}
\newcommand{\lolEB}{{lowlEB}}
\newcommand{\xpol}{{\tt Xpol}}

\newcommand{\NPIPE}{{\tt NPIPE}}
\newcommand{\nside}{\ifmmode {N_{\rm side}} \else $N_{\rm side}$ \fi}

\newenvironment{eqs}
{\begin{subequations}\begin{eqnarray}}
{\end{eqnarray}\end{subequations}}

\newcommand{\myurl}[1]{\href{#1}{\tt #1}}

\def\setsymbol#1#2{\expandafter\def\csname #1\endcsname{#2}}
\def\getsymbol#1{\csname #1\endcsname}

\def\Planck{\textit{Planck}}

\newbox\tablebox    \newdimen\tablewidth
\def\leaderfil{\leaders\hbox to 5pt{\hss.\hss}\hfil}
\def\endPlancktablewide{\tablewidth=\textwidth 
    $$\hss\copy\tablebox\hss$$
    \vskip-\lastskip\vskip -2pt}
\def\tablenote#1 #2\par{\begingroup \parindent=0.8em
    \abovedisplayshortskip=0pt\belowdisplayshortskip=0pt
    \noindent
    $$\hss\vbox{\hsize\tablewidth \hangindent=\parindent \hangafter=1 \noindent
    \hbox to \parindent{$^#1$\hss}\strut#2\strut\par}\hss$$
    \endgroup}
\def\doubleline{\vskip 3pt\hrule \vskip 1.5pt \hrule \vskip 5pt}

\def\L2{\ifmmode L_2\else $L_2$\fi}

\def\DeltaT{\ifmmode \Delta T\else $\Delta T$\fi}
\def\deltat{\ifmmode \Delta t\else $\Delta t$\fi}
\def\fknee{\ifmmode f_{\rm knee}\else $f_{\rm knee}$\fi}
\def\Fmax{\ifmmode F_{\rm max}\else $F_{\rm max}$\fi}
\def\solar{\ifmmode{\rm M}_{\mathord\odot}\else${\rm M}_{\mathord\odot}$\fi}
\def\Msolar{\ifmmode{\rm M}_{\mathord\odot}\else${\rm M}_{\mathord\odot}$\fi}
\def\Lsolar{\ifmmode{\rm L}_{\mathord\odot}\else${\rm L}_{\mathord\odot}$\fi}
\def\inv{\ifmmode^{-1}\else$^{-1}$\fi}
\def\mo{\ifmmode^{-1}\else$^{-1}$\fi}
\def\sup#1{\ifmmode ^{\rm #1}\else $^{\rm #1}$\fi}
\def\expo#1{\ifmmode \times 10^{#1}\else $\times 10^{#1}$\fi}
\def\,{\thinspace}
\def\lsim{\mathrel{\raise .4ex\hbox{\rlap{$<$}\lower 1.2ex\hbox{$\sim$}}}}
\def\gsim{\mathrel{\raise .4ex\hbox{\rlap{$>$}\lower 1.2ex\hbox{$\sim$}}}}

\def\simprop{\mathrel{\raise .4ex\hbox{\rlap{$\propto$}\lower 1.2ex\hbox{$\sim$}}}}
\def\deg{\ifmmode^\circ\else$^\circ$\fi}
\def\pdeg{\ifmmode $\setbox0=\hbox{$^{\circ}$}\rlap{\hskip.11\wd0 .}$^{\circ}
          \else \setbox0=\hbox{$^{\circ}$}\rlap{\hskip.11\wd0 .}$^{\circ}$\fi}
\def\arcs{\ifmmode {^{\scriptstyle\prime\prime}}
          \else $^{\scriptstyle\prime\prime}$\fi}
\def\arcm{\ifmmode {^{\scriptstyle\prime}}
          \else $^{\scriptstyle\prime}$\fi}
\newdimen\sa  \newdimen\sb
\def\parcs{\sa=.07em \sb=.03em
     \ifmmode \hbox{\rlap{.}}^{\scriptstyle\prime\kern -\sb\prime}\hbox{\kern -\sa}
     \else \rlap{.}$^{\scriptstyle\prime\kern -\sb\prime}$\kern -\sa\fi}
\def\parcm{\sa=.08em \sb=.03em
     \ifmmode \hbox{\rlap{.}\kern\sa}^{\scriptstyle\prime}\hbox{\kern-\sb}
     \else \rlap{.}\kern\sa$^{\scriptstyle\prime}$\kern-\sb\fi}
\def\ra[#1 #2 #3.#4]{#1\sup{h}#2\sup{m}#3\sup{s}\llap.#4}
\def\dec[#1 #2 #3.#4]{#1\deg#2\arcm#3\arcs\llap.#4}
\def\deco[#1 #2 #3]{#1\deg#2\arcm#3\arcs}
\def\rra[#1 #2]{#1\sup{h}#2\sup{m}}

\def\dots{\relax\ifmmode \ldots\else $\ldots$\fi}
\def\WHzsr{\ifmmode $W\,Hz\mo\,sr\mo$\else W\,Hz\mo\,sr\mo\fi}
\def\mHz{\ifmmode $\,mHz$\else \,mHz\fi}
\def\GHz{\ifmmode $\,GHz$\else \,GHz\fi}
\def\mKs{\ifmmode $\,mK\,s$^{1/2}\else \,mK\,s$^{1/2}$\fi}
\def\muKs{\ifmmode \,\mu$K\,s$^{1/2}\else \,$\mu$K\,s$^{1/2}$\fi}
\def\muKRJs{\ifmmode \,\mu$K$_{\rm RJ}$\,s$^{1/2}\else \,$\mu$K$_{\rm RJ}$\,s$^{1/2}$\fi}
\def\muKHz{\ifmmode \,\mu$K\,Hz$^{-1/2}\else \,$\mu$K\,Hz$^{-1/2}$\fi}
\def\MJysr{\ifmmode \,$MJy\,sr\mo$\else \,MJy\,sr\mo\fi}
\def\MJysrmK{\ifmmode \,$MJy\,sr\mo$\,mK$_{\rm CMB}\mo\else \,MJy\,sr\mo\,mK$_{\rm CMB}\mo$\fi}
\def\microns{\ifmmode \,\mu$m$\else \,$\mu$m\fi}

\def\muK{\ifmmode \,\mu$K$\else \,$\mu$\hbox{K}\fi}
\def\microK{\ifmmode \,\mu$K$\else \,$\mu$\hbox{K}\fi}
\def\muW{\ifmmode \,\mu$W$\else \,$\mu$\hbox{W}\fi}
\def\kms{\ifmmode $\,km\,s$^{-1}\else \,km\,s$^{-1}$\fi}
\def\kmsMpc{\ifmmode $\,\kms\,Mpc\mo$\else \,\kms\,Mpc\mo\fi}
\providecommand{\sorthelp}[1]{}

\begin{document}


\title{\Planck\ constraints on the tensor-to-scalar ratio}

\author{
M.~Tristram\inst{1}
\and
A.~J.~Banday\inst{2, 3}
\and
K.~M.~G\'{o}rski\inst{4, 5}
\and
R.~Keskitalo\inst{6,7}
\and
C.~R.~Lawrence\inst{4}
\and
K.~J.~Andersen\inst{8}
\and
R.~B.~Barreiro\inst{9}
\and
J.~Borrill\inst{6,7}
\and
H.~K.~Eriksen\inst{8}
\and
R.~Fernandez-Cobos\inst{9}
\and
T.~S.~Kisner\inst{6,7}
\and
E.~Mart\'{\i}nez-Gonz\'{a}lez\inst{9}
\and
B.~Partridge\inst{10}
\and
D.~Scott\inst{11}
\and
T.~L.~Svalheim\inst{8}
\and
H.~Thommesen\inst{8}
\and
I.~K.~Wehus\inst{8}
}

\institute{
Universit\'{e} Paris-Saclay, CNRS/IN2P3, IJCLab, 91405 Orsay, France
\and
Universit\'{e} de Toulouse, UPS-OMP, IRAP, F-31028 Toulouse cedex 4, France
\and
CNRS, IRAP, 9 Av. colonel Roche, BP 44346, F-31028 Toulouse cedex 4, France
\and
Jet Propulsion Laboratory, California Institute of Technology, 4800 Oak Grove Drive, Pasadena, California, U.S.A.
\and
Warsaw University Observatory, Aleje Ujazdowskie 4, 00-478 Warszawa, Poland
\and
Computational Cosmology Center, Lawrence Berkeley National Laboratory, Berkeley, California, U.S.A.
\and
Space Sciences Laboratory, University of California, Berkeley, California, U.S.A.
\and
Institute of Theoretical Astrophysics, University of Oslo, Blindern, Oslo, Norway
\and
Instituto de F\'{\i}sica de Cantabria (CSIC-Universidad de Cantabria), Avda. de los Castros s/n, Santander, Spain
\and
Haverford College Astronomy Department, 370 Lancaster Avenue, Haverford, Pennsylvania, U.S.A.
\and
Department of Physics \& Astronomy, University of British Columbia, 6224 Agricultural Road, Vancouver, British Columbia, Canada
}

\abstract{We present constraints on the tensor-to-scalar ratio $r$ using \Planck\ data. We use the latest release of \planck\ maps (PR4), processed with the \NPIPE\ code, which produces calibrated frequency maps in temperature and polarization for all \planck\ channels from 30\GHz\ to 857\GHz\ using the same pipeline. We computed constraints on $r$ using the $BB$ angular power spectrum, and we also discuss constraints coming from the $TT$ spectrum. Given \planck's noise level, the $TT$ spectrum gives constraints on $r$ that are cosmic-variance limited (with $\sigma_r = 0.093$), but we show that the marginalized posterior peaks towards negative values of $r$ at about the 1.2$\,\sigma$ level. We derived \Planck\ constraints using the $BB$ power spectrum at both large angular scales (the `reionization bump') and intermediate angular scales (the `recombination bump') from $\ell$ = 2 to 150 and find a stronger constraint than that from $TT$, with $\sigma_r = 0.069$. The \planck\ $BB$ spectrum shows no systematic bias and is compatible with zero, given both the statistical noise and the systematic uncertainties. The likelihood analysis using $B$ modes yields the constraint $r < 0.158$ at 95\,\% confidence using more than 50\,\% of the sky. This upper limit tightens to $r < 0.069$ when \planck\ $EE$, $BB$, and $EB$ power spectra are combined consistently, and it tightens further to $r < 0.056$ when the \Planck\ $TT$ power spectrum is included in the combination.  Finally, combining \Planck\ with BICEP2/Keck 2015 data yields an upper limit of $r < 0.044$.}


\date{\today}

\maketitle


\section{Introduction}

While the idea of cosmic inflation was introduced about 40~years ago to solve inherent problems with the canonical hot big-bang model \citep{Brout77, Starobinsky80, Kazanas80, Sato81,Guth80, Linde81, Albrecht82, Linde83}, attention quickly focused on using it as a means to generate cosmological perturbations from quantum fluctuations \citep{Mukhanov81,Mukhanov82,Hawking82,Guth82,Starobinsky82,Bardeen83,Mukhanov85}.  These perturbations include a tensor component (i.e., gravitational waves) as well as the scalar component (i.e., density variations). Inflationary gravitational waves entering the horizon between the epoch of recombination and the present day generate a tensor contribution to the large-scale cosmic microwave background (CMB) anisotropy. Hence, primordial tensor fluctuations contribute to the CMB anisotropies, both in temperature ($T$) and in polarization ($E$ and $B$ modes; \citealt{Seljak97a,Kamionkowski97,Seljak97b}).

As described in \citet{planck2016-l06} and \citet{,planck2016-l10}, the comoving wavenumbers of tensor modes probed by the CMB temperature anisotropy power spectrum have $k \la 0.008\,\Mpc^{-1}$, with very little sensitivity to higher wavenumbers because gravitational waves decay on sub-horizon scales. The corresponding multipoles in the harmonic domain are $\ell \la 100$, for which the scalar perturbations dominate with respect to tensor modes in temperature. The tensor component can be fitted together with the scalar one, and the precision of the \planck\ constraint is limited by the cosmic variance of the large-scale anisotropies.

In polarization, the $EE$ and $TE$ spectra also contain a tensor signal coming from the last-scattering and reionization epochs. The $BB$ power spectrum, however, is treated differently when determining the tensor contribution, since the model does not predict any primordial scalar fluctuations in $BB$.  As a consequence, a primordial $B$-mode signal would be a direct signature of tensor modes. However, depending on the amplitude of the tensor-to-scalar ratio, such a signal may be masked by $E$-mode power that is transformed to $B$-mode power through lensing by gravitational potentials along the line of sight \citep[so-called `$BB$ lensing,'][]{Zaldarriaga98}.  $BB$ lensing has been measured with high accuracy by \planck\ in both harmonic \citep{planck2016-l08} and map \citep{planck2015-XLI} domains, as well as by ground-based observatories POLARBEAR \citep{polarbear17}, SPTpol \citep{Sayre20}, and ACTPol \citep{Choi20}. But a primordial $BB$ tensor signal has not been detected yet.

\begin{figure}[!htbp]
  \centering
  \includegraphics[width=\columnwidth]{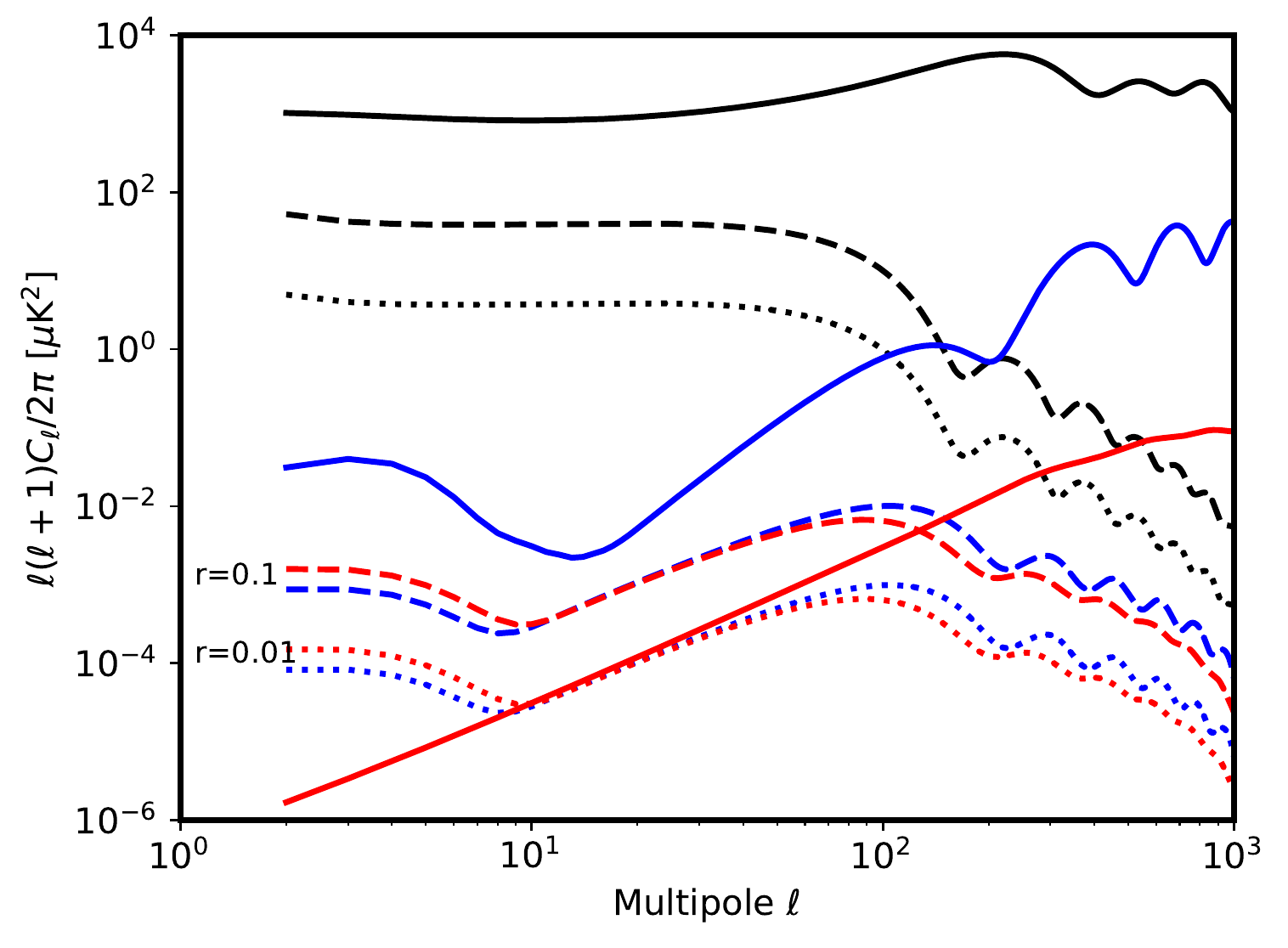}
  \caption{Scalar (thick solid lines) versus tensor spectra for $r=0.1$ (dashed lines) and $r=0.01$ (dotted lines). Spectra for $TT$ are in black, $EE$ in blue, and $BB$ in red.  The red solid line corresponds to the signal from $BB$ lensing.}
  \label{fig:cl_tensor}
\end{figure}

The scalar and tensor CMB angular power spectra are plotted in Fig.~\ref{fig:cl_tensor} for the \planck\ 2018 cosmology and for two values of the tensor-to-scalar ratio, namely $r = 0.1$ and $r=0.01$.  For a further discussion of the tensor-to-scalar ratio and its implications for inflationary models, see \citet{planck2013-p17}, \citet{planck2014-a24}, and \citet{planck2016-l10}. We note that the signal from tensor modes in $EE$ is similar to that in $BB$ modes, which makes $EE$ (in particular at low multipoles) an important data set for tensor constraints. Indeed, limits set by cosmic variance alone for full-sky spectra are $\sigma_r(TT)=0.072$, $\sigma_r(EE)=0.023$, and $\sigma_r(BB)=0.000057$ for $r=0$.
In this paper, we make use of a polarized $E$-$B$ likelihood, which consistently includes the correlated polarization fields $E$ and $B$, and covers the range of multipoles where tensor modes can be constrained using \planck\ data (i.e., from $\ell=2$ to $\ell=150$).

At present the tightest $B$-mode constraints on $r$ come from the BICEP/Keck measurements (BK15; \citealt{Bicep2018limit}), which cover approximately $400\,{\rm deg}^2$ centred on ${\rm RA}=0^{\rm h}$, ${\rm Dec}=-57\pdeg5$.  These measurements probe the peak of the $B$-mode power spectrum at around $\ell=100$, corresponding to gravitational waves with $k\approx 0.01\,\Mpc^{-1}$ that enter the horizon during recombination (i.e., somewhat smaller than the scales contributing to the \Planck\ temperature constraints on $r$). The results of BK15 give a limit of $r < 0.07$ at 95\,\% confidence, which tightens to $r < 0.06$ in combination with \planck\ temperature and other data sets.

\citet{planck2016-l05} presented \Planck\ $B$-mode constraints from the 100- and 143-GHz HFI channels with a 95\,\% upper limit of $r < 0.41$ (at a somewhat larger pivot scale, as described in the next section), using only a limited number of multipoles around the so-called `reionization bump' ($2\leq\ell\leq29$). 
Using \Planck\ \NPIPE\ maps \citep{planck2020-LVII}, called \Planck\ Release 4 (PR4), we are now able to constrain the $BB$ power spectrum for a much larger number of modes, including both the reionization bump at large angular scales ($\ell \la 30$) and the so-called `recombination bump' at intermediate scales ($50\la\ell\la 150$).
In this paper, we first describe, in Sect.~\ref{sec:model}, the cosmological model used throughout the analysis. We then detail the data and the likelihoods in Sect.~\ref{sec:data_lik}. Section~\ref{sec:tt} focuses on constraints from $TT$ and in particular the impact of the \lowl\ data in temperature. Section~\ref{sec:bb} gives constraints from the $BB$ angular power spectrum using \planck\ data, while results from the full set of polarization power spectra are given in Sect.~\ref{sec:pol}. In Sect.~\ref{sec:combined}, we combine all data sets to provide the most robust constraints on $BB$ coming from \planck\ and in combination with other CMB data sets, such as the results from the BICEP/Keck Collaboration. Finally, we provide details of several parts of our analysis in a set of appendices, specifically describing the transfer function for $BB$, the \hillipop\ likelihood, large-scale polarized power spectra, the cross-spectrum correlation matrix, comparison between PR3 and PR4, robustness tests, triangle plots for \LCDM+r parameters, and comparison with other $BB$ spectrum measurements.

\section{Cosmological model}
\label{sec:model}
We use the base-\LCDM\ model, which has been established over the last couple of decades to be the simplest viable cosmological model, in particular with the \planck\ results \citep[e.g.,][]{planck2016-l06}.
In this model, we assume purely adiabatic, nearly scale-invariant perturbations at very early times, with curvature-mode (scalar) and tensor-mode power spectra parameterized by
\begin{eqnarray}
	\mathcal{P}_{\rm s}(k) &=& A_{\rm s} \left(\frac{k}{k_0}\right)^{n_{\rm s}-1},		\label{eq:Ps} \\
	\mathcal{P}_{\rm t}(k) &=& A_{\rm t} \left(\frac{k}{k_0}\right)^{n_{\rm t}},		\label{eq:Pt} 
\end{eqnarray}
where $A_{\rm s}$ and $A_{\rm t}$ are the initial super-horizon amplitudes for curvature and tensor perturbations, respectively. The primordial spectral indexes for scalar ($n_{\rm s}$) and tensor ($n_{\rm t}$) perturbations are taken to be constant. This means that we assume no `running,' i.e., a pure power-law spectrum with $d n_{\rm s} / d\ln k = 0$. 
We set the pivot scale at $k_0 = 0.05\,\Mpc^{-1}$, which roughly corresponds to approximately the middle of the logarithmic range of scales probed by \planck;
with this choice, $n_{\rm s}$ is not strongly degenerate with the amplitude parameter $A_{\rm s}$.
Note that for historical reasons, the definitions of $n_{\rm s}$ and $n_{\rm t}$ differ, so that a scale-invariant scalar spectrum corresponds to $n_{\rm s} = 1$, while a scale-invariant tensor spectrum corresponds to $n_{\rm t} = 0$. 

The late-time parameters, on the other hand, determine the linear evolution of perturbations after they re-enter the Hubble radius. We use the basis ($\Omega_{\rm b}h^2$, $\Omega_{\rm c}h^2$, $\theta_{\ast}$, $\tau$) following the approach in \planck\ cosmological studies \citep{planck2016-l06}, where $\Omega_{\rm b}h^2$ is the baryon density today, $\Omega_{\rm c}h^2$ is the cold dark matter density today, $\theta_{\ast}$ is the observed angular size of the sound horizon at recombination, and $\tau$ is the reionization optical depth.

The amplitude of the small-scale linear CMB power spectrum is proportional to $A_{\rm s}e^{-2\tau}$. Because \planck\ measures this amplitude very accurately, there is a tight linear constraint between $\tau$ and $\ln A_{\rm s}$.
For this reason, we usually adopt $\ln A_{\rm s}$ as a base parameter with a flat prior; $\ln A_{\rm s}$ has a significantly more Gaussian posterior than $A_{\rm s}$.  A linear parameter redefinition then allows the degeneracy between $\tau$ and $A_{\rm s}$ to be explored efficiently.  Note that the degeneracy between $\tau$ and $A_{\rm s}$ is broken by the relative amplitudes of large-scale temperature and polarization CMB anisotropies and by the effect of CMB lensing.

We define $r \equiv A_{\rm t}/A_{\rm s}$, the primordial tensor-to-scalar ratio defined explicitly at the scale $k_0=0.05,\Mpc^{-1}$. Our constraints are only weakly sensitive to the tensor spectral index, $n_{\rm t}$. We adopt the single-field-inflation consistency relation $n_{\rm t}=-r/8$. Note that the Planck Collaboration also discussed $r$ constraints for $k_0=0.002,\Mpc^{-1}$ \citep{planck2016-l05}. Given the definitions in Eqs.~\eqref{eq:Ps} and \eqref{eq:Pt}, the tensor-to-scalar ration scales with $(0.05/0.002)^{-r/8}$, which means that $r_{0.002}$ is lower by 4\,\% at $r\simeq0.1$ compared to $r_{0.05}$ and less than 0.4\,\% lower for $r<0.01$.

In this work, we use an effective tensor-to-scalar ratio $r_{\rm eff}$, which we extend into the negative domain by modifying the Boltzmann-solver code \texttt{CLASS}~\citep{Blas2011}.  While negative tensor amplitudes are unphysical, this approach will allow us to derive posteriors without boundaries, facilitating detection of potential biases, and enabling us to determine a more accurate statistical definition of the constraints on $r$. With $r_{\rm eff}$ we are able to independently discuss both the uncertainty of $r$ ($\sigma_r$) and corresponding upper limits (depending on the maximum a posteriori probability). 
In the rest of this paper, we simply write $r$ as the effective tensor-to-scalar ratio, and report upper limits for positive tensor amplitudes, for which $r_{\rm eff} = r$. 
We use 95\,\% confidence levels when reporting upper limits, and a 68\,\% confidence interval with the maximum a posteriori probability.

\section{Data and likelihoods}
\label{sec:data_lik}

\subsection{Data and simulations \label{sec:data}}

The sky measurements used in this analysis are the PR4 maps available from the Planck Legacy Archive\footnote{\myurl{https://pla.esac.esa.int}} (PLA) and from the National Energy Research Scientific Computing Center (NERSC).\footnote{\myurl{https://portal.nersc.gov/project/cmb/planck2020}} They have been produced with the \NPIPE\ processing pipeline, which creates calibrated frequency maps in temperature and polarization from the \Planck\ Low Frequency Instrument (LFI) and High Frequency Instrument (HFI) data. As described in \citet{planck2020-LVII}, \NPIPE\ processing includes several improvements, resulting in lower levels of noise and systematics in both frequency and component-separated maps at essentially all angular scales, as well as notably improved internal consistency between the various frequencies.

\NPIPE\ achieves an overall lower noise level in part by incorporating the data acquired during the 4-minute spacecraft repointing manoeuvres that take place between the 30-to-70-min stable science scans. Residual systematics are suppressed using a mitigation strategy that combines aspects of both LFI and HFI processing pipelines. Most importantly, gain fluctuations, bandpass mismatch, and other systematics are formulated into time-domain templates that are fitted and subtracted as a part of the mapmaking process.  Degeneracies between sky polarization and systematic templates are broken by constructing a prior of the polarized foreground sky using the extreme polarization-sensitive frequencies (30, 217, and 353\,GHz).

\begin{figure}[htbp!]
	\centering
	\includegraphics[width=\columnwidth]{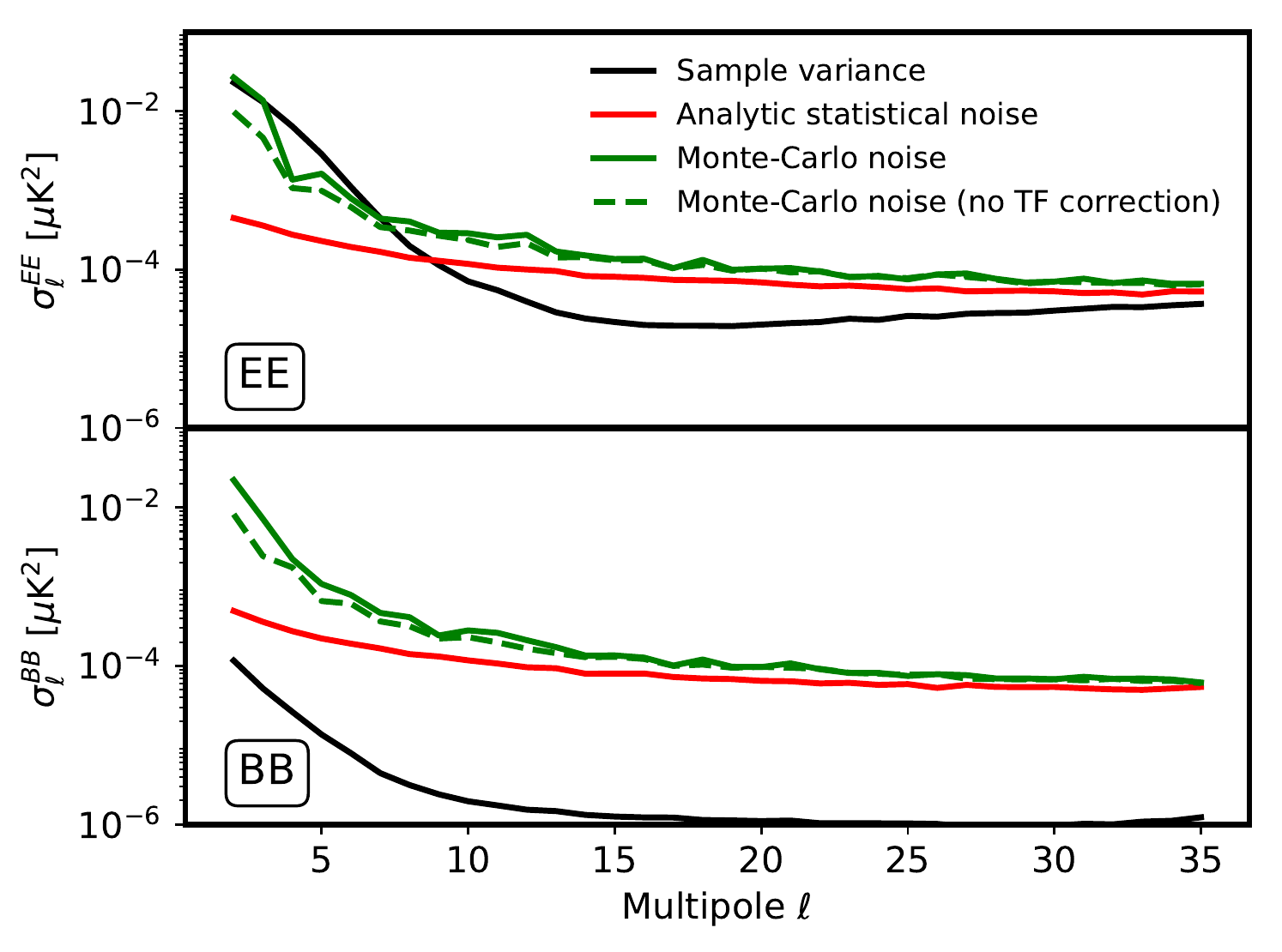}
	\caption{Variances for cross-spectra in $EE$ and $BB$ based on PR4 simulations, including: cosmic (sample) variance (black); analytic statistical noise (red); and PR4 noise from Monte Carlo simulations (green), including noise and systematics with (solid line) or without (dashed line) correction for the transfer function. The sky fraction used here is 80\,\%, as it illustrates well the effect of both systematics and transfer-function corrections \citep[][]{planck2020-LVII}.}
	\label{fig:cl_var}
\end{figure}

Moreover, the PR4 release comes with 400 simulations of signal, noise, and systematics, component-separated into CMB maps, which allow for an accurate characterization of the noise and systematic residuals in the \planck\ maps. This is important because \planck\ polarization data are cosmic-variance-dominated only for a few multipoles at very large scales in $EE$ ($\ell < 8$, as shown in Fig.~\ref{fig:cl_var}). These simulations, even though limited in number, represent a huge effort in terms of CPU time.  They are essential in order to compute the following two additional quantities.\\
First, the end-to-end transfer function from the data reduction (including TOI processing, mapmaking and template fitting for mitigation of systematics, component separation, and power-spectrum estimation). The transfer function is defined as the ratio between the estimated output power spectrum and the input one, averaged over all the simulations \citep[see section~4.3 of][for the details]{planck2020-LVII}.\\
Second, the covariance of the data (here we use the cross-power spectra), which is the only way to propagate uncertainties when those are dominated by systematics (from the instrument or from foregrounds).\\
Note that these two quantities estimated from the simulations are directly related to two different characteristics of the final parameter posteriors: the bias of the mean (the transfer function); and the width of the posterior (as propagated into parameter constraints by the covariance matrix in the likelihood).  They can be separated from each other, meaning that one systematic effect can easily produce a significant bias without any strong impact on the variance, while another effect can produce a large increase of the variance with no associated bias.

The \NPIPE\ simulations include the systematic effects relevant for polarization studies, specifically analogue-to-digital-converter nonlinearities, gain fluctuations, bandpass mismatch between detectors, correlated noise (including 4-K line residuals), and full-beam convolutions for each detector.

The use of a polarization prior in \NPIPE\ processing causes a suppression of large-scale ($\ell\,{<}\,20$) CMB polarization, which needs to be corrected. 
As explained in \citet{planck2020-LVII}, allowing for a non-trivial transfer function is a compromise between measuring very noisy but unbiased large-scale polarization from all low-$\ell$ modes, and filtering out the modes that are most affected by the calibration uncertainties left in the data by the \Planck\ scan strategy. 
As detailed in \citet{planck2020-LVII}, the transfer function to correct for this bias is determined from simulations. It is then used to correct the power spectrum estimates, just as instrumental beam and pixel effects must be deconvolved. 
Due to the fact that $E$ modes dominate the CMB polarization, the simulations do not yield a definitive measurement of the $B$-mode transfer function. We have chosen to conservatively deconvolve the $E$-mode transfer function from the $B$-mode spectrum in order to provide a robust upper limit on the true $B$-mode spectrum. 
Indeed, when regressing the templates fitted during the map-making process with pure $E$ and $B$ CMB maps, we found a similar impact on the $EE$ and $BB$ power spectra (see Appendix~\ref{ann:bbtf}). Moreover, in the situation where primordial $B$-mode power is not detected, the transfer function correction essentially increases the variance estimate at low multipoles, which propagates the uncertainty induced by the degeneracy between the sky and the systematic templates used in \NPIPE. Note that this uncertainty is small compared to the impact of systematics in the error budget (see Fig.~\ref{fig:cl_var}).

To compute unbiased estimates of the angular power spectra, we perform cross-correlations of two independent splits of the data. As shown in \citet{planck2020-LVII}, the most appropriate split for the \planck\ data is represented by the detector-set (hereafter `detset') maps, comprising two subsets of maps at each frequency, with nearly independent noise characteristics, made by combining half of the detectors. This was obtained by processing each split independently, in contrast to the detset maps produced in the previous \Planck\ releases. Note that time-split maps (made from, e.g., `odd-even rings' or `half-mission data') share the same instrumental detectors, and therefore exhibit noise correlations due to identical spectral bandpasses and optical responses. The use of time-split maps is subject to systematic biases in the cross-power spectra \citep[see section~3.3.3 in][]{planck2016-l05}, as well as underestimation of the noise properties in computing the half-differences (which must be compensated by a rescaling of the noise in the PR3 as described in Appendix~A.7 of \citealt{planck2016-l03}). Hence we use detset splits here.

Uncertainties at the power-spectrum level are dominated by noise and systematics, as illustrated in Fig.~\ref{fig:cl_var}. Thanks to the \NPIPE\ processing, we are now able to show the impact of the systematics at low $\ell$. This is illustrated by comparing the PR4 end-to-end noise (based on the Monte Carlo simulations, including instrumental noise, systematics, and foreground uncertainties, and corrected for the transfer function both in $EE$ and $BB$) with the propagation of the statistical noise coming from the analytic pixel-pixel covariance matrix. The systematic uncertainties dominate at $\ell \la 15$, then slowly decrease so that the effective uncertainties converge towards the analytic estimate at higher multipoles.

\subsection{Polarized sky masks \label{sec:masks}}

\begin{figure}[htbp!]
	\centering
	\includegraphics[width=\columnwidth]{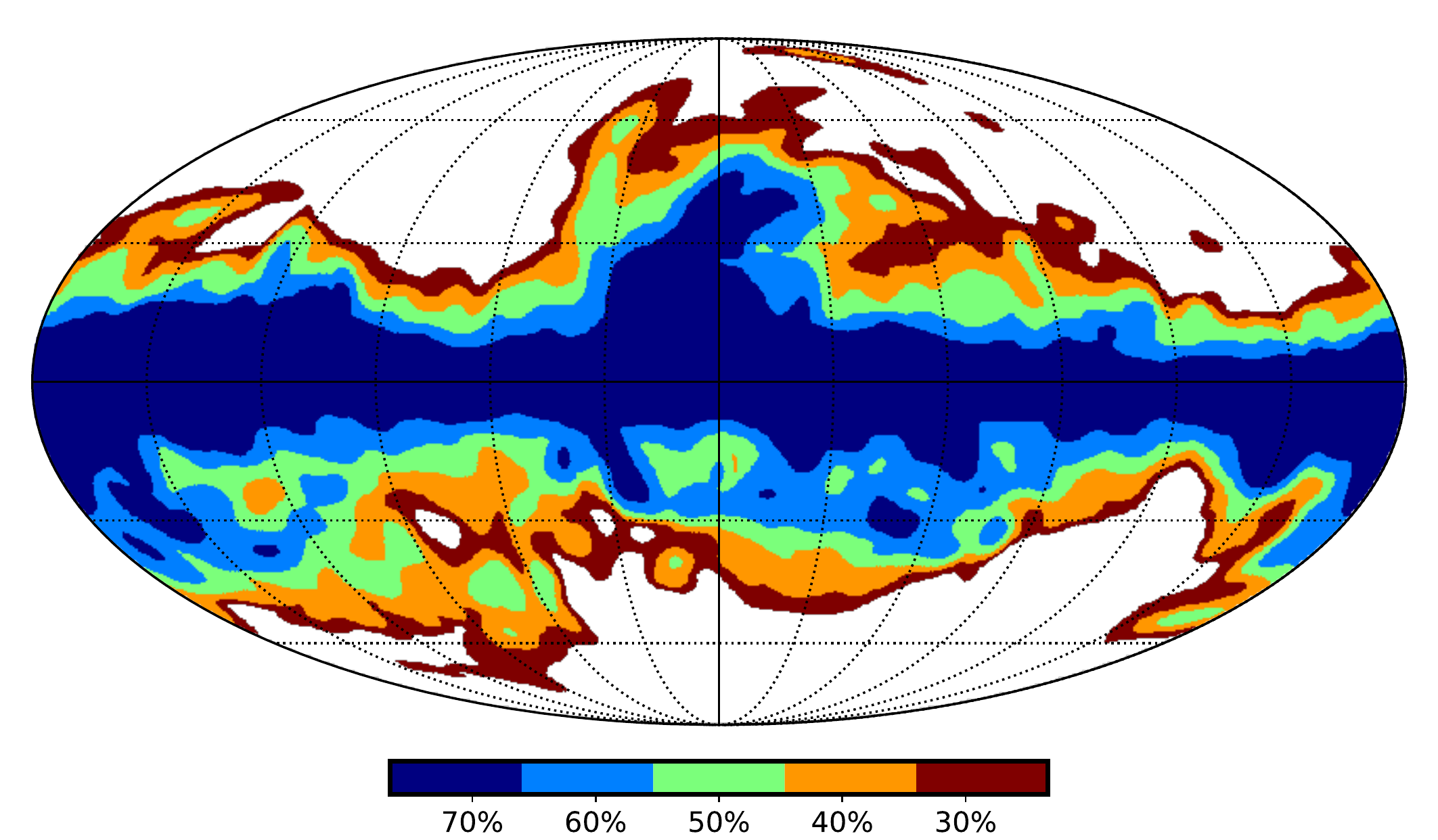}
	\caption{Galactic masks used for the \planck\ likelihoods. The mask shown in dark blue indicates the sky rejected in order to retain a 70\,\% sky fraction for analysis. The masks shown in light blue, green, orange and red incrementally omit further parts of the sky, corresponding in turn to 60, 50, 40 and 30\,\% retained sky fractions, the latter shown in white.}
	\label{fig:masks}
\end{figure}

Foreground residuals in the foreground-cleaned maps dominate the polarized CMB signal near the Galactic plane. 
To avoid contamination from these residuals in the cosmological analysis, we mask the Galactic plane. We use a series of different retained sky fractions (from 30\,\% to 70\,\%) to check the consistency of our results with respect to foreground residuals (Fig.~\ref{fig:masks}).

The masks used in this analysis are a combination of a mask for polarization intensity (to avoid polarized foreground residuals), a mask for total intensity (to avoid potential temperature-to-polarization leakage residuals), and the confidence mask for component separation provided by the \Planck\ Colaboration.  The intensity mask is obtained by thresholding the combination of the 353-GHz intensity map (which traces dust) scaled to 143\,GHz, and the 30-GHz intensity map (which traces synchrotron) scaled to 100\,GHz.  The polarization map is constructed similarly. Both foreground tracers are smoothed beforehand with a 10\deg\ Gaussian window function.

The impact of the emission of extragalactic polarized sources on the power spectra is negligible, given the \planck\ resolution and noise level. The confidence mask for component separation ensures the masking of the strongest sources, which could also produce residuals through temperature-to-polarization leakage.

\subsection{Likelihoods}

Table~\ref{tab:lik} summarizes the likelihoods used in this analysis, which are described below.
\begin{table*}[htbp!] 
 \begingroup
\caption{Summary of the likelihoods used in this paper.} 
\label{tab:lik}
\nointerlineskip
\vskip -3mm
\setbox\tablebox=\vbox{
    \newdimen\digitwidth
    \setbox0=\hbox{\rm 0}
    \digitwidth=\wd0
    \catcode`*=\active
    \def*{\kern\digitwidth}
    \newdimen\signwidth
    \setbox0=\hbox{+}
    \signwidth=\wd0
    \catcode`!=\active
    \def!{\kern\signwidth}
\halign{\hbox to 0.9in{#\leaderfil}\tabskip=2.0em&
    \hfil#\hfil\tabskip=2em&
    \hfil#\hfil\tabskip=2em&
    \hfil#\hfil\tabskip=2em&
    #\hfil\tabskip=0pt\cr
\noalign{\doubleline}
\omit\hfil Name\hfil&Mode&$\ell$ range&\Planck\ release&\omit\hfil Description\hfil\cr
\noalign{\vskip 3pt\hrule\vskip 4pt}
\lowT$^{\rm a}$& TT& *2--30**& PR3& \commander\ likelihood for Temperature\cr
\hlp TT$^{\rm b}$& TT& 30--2500& PR4& \hillipop\ likelihood for \highl\ TT\cr
\hlp TTTE$^{\rm b}$& TT+TE& 30--2500& PR4& \hillipop\ likelihood for \highl\ TT+TE\cr
\noalign{\vskip 5pt}
\lolE$^{\rm b}$& EE& *2--150*& PR4& \lollipop\ likelihood for \lowl\ EE\cr
\lolB$^{\rm b}$& BB& *2--150*& PR4& \lollipop\ likelihood for \lowl\ BB\cr
\lolEB$^{\rm b}$& EE+BB+EB& *2--150*& PR4& \lollipop\ likelihood for \lowl\ EE+BB+EB\cr
\noalign{\vskip 4pt\hrule\vskip 3pt}}}
\endPlancktablewide
\tablenote{{\rm a}} {\tiny available from \myurl{https://pla.esac.esa.int}}\par
\tablenote{{\rm b}} {\tiny available from \myurl{https://github.com/planck-npipe}}\par
\endgroup
\end{table*}

\subsubsection{Low-$\ell$ temperature likelihood}
We use the \Planck\ public \lowl\ temperature-only likelihood based on the PR3 CMB map recovered from the component-separation procedure (specifically {\tt Commander}) described in detail in \citet{planck2016-l05}.  At large angular scales, \planck\ temperature maps are strongly signal-dominated, and there is no expected gain in updating this likelihood with the PR4 data.

As discussed in \cite{planck2014-a24}, the \lowl\ temperature data from \planck\ have a strong impact on the $r$ posterior and the derivation of the corresponding constraints. This is because the deficit of power in the measured $C_\ell$s at \lowl\ in temperature (see the discussions in \citealt{planck2013-p11} and \citealt{planck2016-LI}) lowers the probability of tensor models, which `add' power at low multipoles.  This shifts the maximum in the posterior of $r$ towards low values (or even negative values when using $r_{\rm eff}$, as we show in Sect.~\ref{sec:tt}).

\subsubsection{High-$\ell$ likelihood}
\label{sec:lik_highl}

At small angular scales ($\ell > 30$), we use the \hillipop\ likelihood, which can include the $TT$, $TE$, and/or $EE$ power spectra. \hillipop\ has been used as an alternative to the public \planck\ likelihood in the 2013 and 2015 \planck\ releases \citep{planck2013-p08,planck2014-a13}, and is described in detail in~\citet{couchot2017}.  In this paper, the \hillipop\ likelihood is applied to the PR4 detset maps at 100, 143, and 217\GHz.  We focus on the $TT$ spectra, since there is marginal additional information at small scales in $TE$ or $EE$ for tensor modes, due to \Planck\ noise. We only make use of $TE$ in Sect.~\ref{sec:combined} in order to help constrain the spectral index $n_{\rm s}$.
The likelihood is a spectrum-based Gaussian approximation, with semi-analytic estimates of the $C_\ell$ covariance matrix based on the data. The cross-spectra are debiased from the effects of the mask and the beam leakage using \xpol\ (a generalization to polarization of the algorithm presented in \citealt{tristram2005}\footnote{\myurl{https://gitlab.in2p3.fr/tristram/Xpol}}) before being compared to the model, which includes CMB and foreground residuals.  The beam window functions are evaluated using {\sc QuickPol} \citep{hivon17}, adapted to the PR4 data.  These adaptations include an evaluation of the beam-leakage effect, which couples temperature and polarization modes due to the beam mismatch between individual detectors.

The model consists of a linear combination of the CMB power spectrum and several foregrounds residuals. These are:
\begin{itemize}
\item Galactic dust (estimated directly from the 353-GHz channel); 
\item the cosmic infrared background \citep[as measured in][]{planck2013-pip56}; 
\item thermal Sunyaev-Zeldovich emission \citep[based on the \planck\ measurement reported in][]{planck2013-p05b}; 
\item kinetic Sunyaev-Zeldovich emission, including homogeneous and patchy reionization components from \cite{shaw12} and \cite{battaglia13};
\item a tSZ-CIB correlation consistent with both models above; and 
\item unresolved point sources as a Poisson-like power spectrum with two components (extragalactic radio galaxies and infrared dusty galaxies).
\end{itemize}

On top of the cosmological parameters associated with the computation of the CMB spectrum, with \hillipop\ we sample seven foreground amplitudes (one per emission source, the spectral energy density rescaling the amplitude for each cross-frequency being fixed) and six nuisance parameters (one overall calibration factor plus intercalibrations for each map). See Appendix~\ref{ann:hillipop} for more details.

\subsubsection{Large-scale polarized likelihood}
\label{sec:lik:lol}

We construct a polarized $E$-$B$ likelihood based on power spectra, focusing on the large scales where the tensor signal is dominant. Because it carries very little information about the tensor modes, we do not include the $TE$ spectrum in this analysis.

In polarization, especially at large angular scales, foregrounds are stronger relative to the CMB than in temperature, and cleaning the \planck\ frequencies using $C_\ell$ templates in the likelihood (as done in temperature) is not accurate enough. In order to clean sky maps of polarized foregrounds, we use the \commander\ component-separation code \citep{eriksen2008}, with a model that includes three polarized components, namely the CMB, synchrotron, and thermal dust emission. \commander\ was run on each detset map independently, as well as on each realization from the PR4 Monte Carlo simulations. Maps are available on the PLA in {\tt HEALPix}\footnote{\myurl{http://healpix.sourceforge.net}} format~\citep{gorski2005} at a resolution $\nside=2048$.

To compute unbiased estimates of the angular power spectra, we calculate the cross-correlation of the two detset maps. We make use of two different angular cross-power spectra estimators (described below), which are then concatenated to produce a full-multipole-range power spectrum. There is no information loss in this process, since the covariances are deduced using Monte Carlo simulations including the correlations over the entire multipole range.
\begin{itemize}
\item For multipoles $2\leq \ell \leq 35$, we compute power spectra using an extension of the quadratic maximum likelihood estimator \citep{tegmark01} adapted for cross-spectra in \citet{vanneste18}.\footnote{\myurl{https://gitlab.in2p3.fr/xQML/xQML}} At multipoles below 40, it has been shown to produce unbiased polarized power spectra with almost optimal errors.  We use downgraded $\nside\,{=}\,16$ maps after convolution with a cosine apodizing kernel $b_\ell = \frac{1}{2}\left\{1+\cos\pi(\ell-1)/(3\nside-1)\right\}$.
The signal is then corrected with the PR4 transfer function, to compensate for the filtering induced by the degeneracies between the signal and the templates for systematics in the mapmaking procedure (see Sect.~\ref{sec:data}).

\item For multipoles $35<\ell<300$, we compute power spectra with a classical pseudo-$C_\ell$ estimator \xpol\ (Sect.~\ref{sec:lik_highl}).  We used $\nside=1024$ maps and the native beam of \commander\ maps (i.e., 5\arcm). In this case, we apodize the mask (see Sect.~\ref{sec:masks}) with a 1\deg\ Gaussian taper. Given the low signal-to-noise ratio in polarization, we bin the spectra with $\Delta\ell = 10$.
\end{itemize}

\begin{figure*}[!htbp]
	\centering
	\includegraphics[width=\columnwidth]{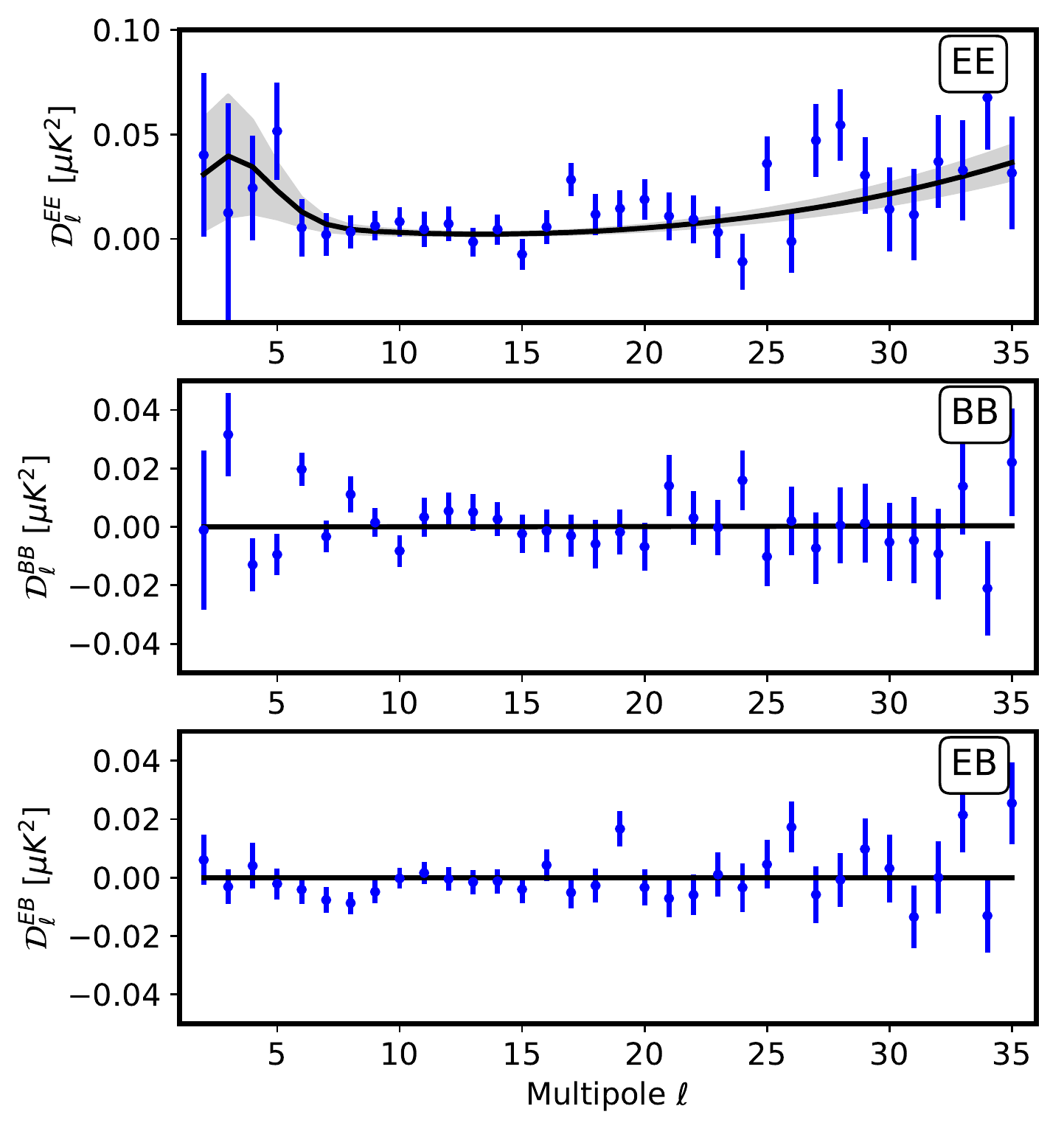}
	\includegraphics[width=\columnwidth]{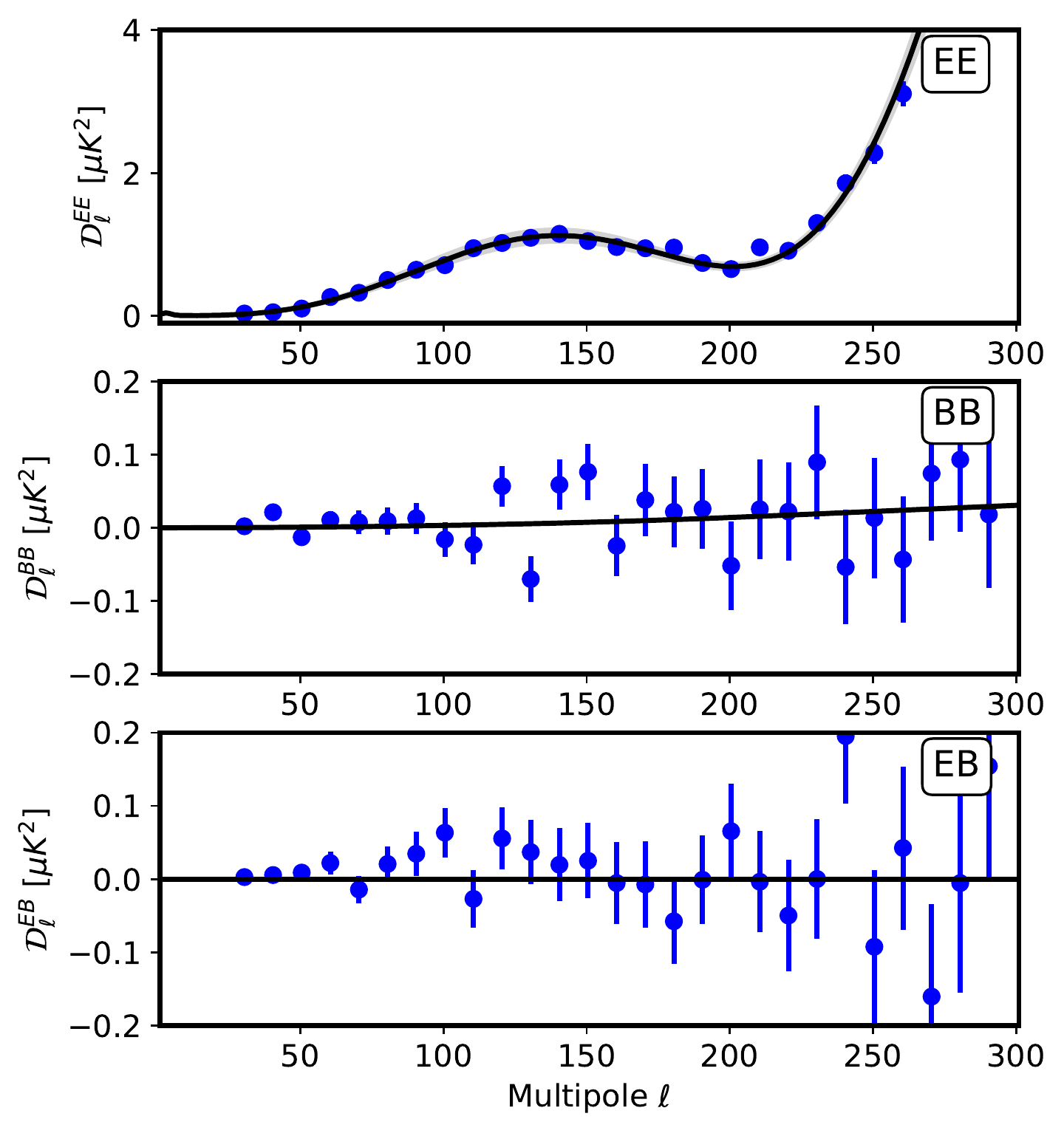}
	\caption{$EE$, $BB$, and $EB$ power spectra of the CMB computed on 50\,\% of the sky with the PR4 maps at low (left panels) and intermediate multipoles (right panels). The \Planck\ 2018 \LCDM\ model is plotted in black. Grey bands represent the associated cosmic variance. Error bars are deduced from the PR4 Monte Carlo simulations.  Correlations between data points are given in Appendix~\ref{ann:corrmat}. A simple $\chi^2$ test shows no significant departure from the model for any of these spectra.}
	\label{fig:cl_EE_BB_EB}
\end{figure*}

The $EE$, $BB$, and $EB$ power spectra estimates are presented in Fig.~\ref{fig:cl_EE_BB_EB} for 50\,\% of the sky, which provides the best combination of sensitivity and freedom from foreground residuals. Power spectra computed on different sky fractions (using masks from Sect.~\ref{sec:masks}) are compared in Fig.~\ref{fig:cl_galcut}. 
A simple $\chi^2$ test on the first 34 multipoles shows no significant departure from the \Planck\ 2018 \LCDM\ model for any of these spectra. The `probability to exceed' values (PTE) for the $EE$, $BB$, and $EB$ spectra on the first 34 multipoles are 0.27, 0.21, and 0.26, respectively.
The most extreme multipole in the $BB$ spectrum is $\ell=6$, which, for a Gaussian distribution, would correspond conditionally to a 3.4$\,\sigma$ outlier (reducing to 2.3$\,\sigma$ after taking into account the look-elsewhere effect, including the first 34 multipoles). However, at such low multipoles, the distribution is not Gaussian and the PTE are certainly higher than the numbers of $\sigma$ would suggest.  In $EE$, the largest deviation from the model is for $\ell=17$ at 3.1$\,\sigma$ and in $EB$ it is $\ell=19$ at 2.7$\,\sigma$.

The $C_\ell$ covariance matrix is computed from the PR4 Monte Carlos.  For each simulation, we compute the power spectra using both estimators.  The statistical distribution of the recovered $C_\ell$ then naturally includes the effect of the components included in the Monte Carlo, namely the CMB signal, instrumental noise, \planck\ systematic effects incorporated in the PR4 simulations (see Sect.~\ref{sec:data}), component-separation uncertainties, and foreground residuals.  The residual power spectra (both for the simulations and the data) are shown in Fig.~\ref{fig:cl_residuals}.

Given the \planck\ noise level in polarization, we focus on multipoles below $\ell=150$, which contain essentially all the information on tensor modes in the \planck\ CMB angular power spectra.  At those scales, and given \planck\ noise levels, the likelihood function needs to consistently take into account the two polarization fields $E$ and $B$, as well as all correlations between multipoles and modes ($EE$, $BB$, and $EB$).

 \lollipop\ (LOw-$\ell$ LIkelihood on POlarized Power-spectra) is a \Planck\ \lowl\ polarization likelihood based on cross-spectra, and was previously applied to \Planck\ $EE$ data for investigating the reionization history in \citet{planck2014-a25}.  The version used here is updated to use cross-spectra calculated on component-separated CMB detset maps processed by \commander\ from the PR4 frequency maps.  Systematic effects are considerably reduced in cross-correlation compared to auto-correlation, and \lollipop\ is based on cross-power spectra for which the bias is zero when the noise is uncorrelated between maps.  It uses the approximation presented in \citet{hamimeche08}, modified as described in \citet{mangilli15} to apply to cross-power spectra. The idea is to apply a change of variable $C_\ell \rightarrow X_\ell$ so that the new variable $X_\ell$ is nearly Gaussian-distributed.  Similarly to \citet{hamimeche08}, we define
\begin{equation}
  X_\ell = \sqrt{ C_\ell^{\rm f} + O_\ell} \,\, g{\left(\frac{\widetilde{C}_\ell + O_\ell}{C_\ell + O_\ell}\right)} \,\, \sqrt{ C_\ell^{\rm f} + O_\ell} ,
\label{eq:xell}
\end{equation}
where $g(x)=\sqrt{2(x-\ln(x)-1)}$, $\widetilde{C}_\ell$ are the measured cross-power spectra, $C_\ell$ are the power spectra of the model to be evaluated, $C_\ell^{\rm f}$ is a fiducial model, and $O_\ell$ are the offsets needed in the case of cross-spectra. 
For multi-dimensional CMB modes (here we restrict ourselves to $E$ and $B$ fields only), the $C_\ell$ generalise to $\tens{C}_\ell$, a $2\times2$ matrix of power spectra,
\begin{equation}
	\tens{C}_\ell = 
	\left(
	\begin{array}{ccc}
		C_\ell^{EE} +O_\ell^{EE} & C_\ell^{EB} \\
		C_\ell^{BE} & C_\ell^{BB} + O_\ell^{BB}
	\end{array}
	\right) \,,
\end{equation}
and the $g$ function is applied to the eigenvalues of $\tens{C}^{-1/2}_\ell \widetilde{\tens{C}}_\ell \tens{C}^{-1/2}_\ell$ (with $\tens{C}^{-1/2}$ the square root of the positive-definite matrix $\tens{C}$). In the case of auto-spectra, the offsets $O_\ell$ are given by the noise bias effectively present in the measured power spectra. For cross-power spectra, the noise bias is zero, and we use effective offsets defined from the $C_\ell$ noise variance:
\begin{equation}
	\Delta C_\ell \equiv \sqrt{ \frac{2}{2\ell+1}} O_\ell .
\end{equation}

The distribution of the new variable $X_\ell \equiv \text{vecp}(\tens{X}_\ell)$, the vector of distinct elements of $\tens{X}_\ell$, can be approximated as Gaussian, with a covariance given by the covariance of the $C_\ell$s.
The likelihood function of the $C_\ell$ given the data $\widetilde{C}_\ell$ is then 
\begin{equation}
  -2\ln P(C_\ell|\widetilde{C}_\ell)=\sum_{\ell \ell'} X^{\sf T}_\ell \tens{M}^{-1}_{\ell \ell'} X_{\ell'}.
\end{equation}
Uncertainties are incorporated into the $C_\ell$-covariance matrix $\tens{M}_{\ell\ell'}$, which is evaluated after applying the same pipeline (including \commander\ component separation and cross-spectrum estimation on each simulation) to the Monte Carlo simulations provided in  PR4. 
While foreground emission and the cleaning procedure are kept fixed in the simulations (so that we cannot include uncertainties arising from an imperfect foreground model), the resulting $C_\ell$ covariance consistently includes CMB sample variance, statistical noise, and systematic residuals, as well as foreground-cleaning uncertainties, together with the correlations induced by masking. These uncertainties are then propagated through the likelihood up to the level of cosmological parameters. Figures of the correlation matrices are given in Appendix~\ref{ann:corrmat}.

Using this approach, we are able to derive three different likelihoods, one using only information from $E$ modes (\lolE), one using only information from $B$ modes (\lolB), and one using $EE$+$BB$+$EB$ spectra (\lolEB).  We have used these likelihoods from $\ell=2$ up to $\ell=300$ with a nominal range of $\ell=[2,150]$, since multipoles above $\ell \simeq 150$ do not contribute to the result due to the \Planck\ noise (see Sect.~\ref{sec:bb}). 

The approach used in this paper is different from the one used for the \planck\ 2018 results. Indeed, in \cite{planck2016-l05}, the probability density of the polarized spectra at low multipoles was modelled with a polynomial function adjusted on simulations in which only $\tau$ is varied, with all other cosmological parameters in a \LCDM\ model fixed to the \planck\ 2018 best-fit values.  As a consequence, the probability density is not proportional to the likelihood $\mathcal{L}(\Omega^{\rm model} | C_\ell^{\rm data})$ when the model is not \LCDM\ (and in particular for our case \LCDM+$r$), and even in the \LCDM\ case it neglects correlations with other parameters that affect the posterior on $\tau$.
In addition, the simulations used in \cite{planck2016-l05} were generated with the same CMB realization for the mapmaking solution. Cosmic variance was included afterwards by adding CMB realizations on top of noise-only maps, neglecting correlations between foregrounds or systematic templates and the CMB. The information in polarization at \lowl\ was then extracted using a polynomial function fitted to the distribution from simulations. While this is supposed to empirically take into account the effects of systematics on the likelihood shape, it does not include $\ell$-by-$\ell$ correlations, and is limited in the $C_\ell$ power that one can test (for example imposing a strong prior on the $EE$ power at $\ell = 3$).  As a consequence, the combination of those two effects reduces the covariance, especially at low multipoles, leading to error bars (especially on $\tau$) that are underestimated.

\section{Constraints from \textit{TT}}
\label{sec:tt}

To derive constraints on the tensor-to-scalar ratio from the temperature power spectrum, we use the \highl\ \hillipop\ likelihood for $30\leq\ell\leq2500$, and the \commander\ likelihood (\lowT) in temperature for $\ell < 30$, with a prior on the reionization optical depth to break the degeneracy with the scalar amplitude $A_{\rm s}$. We use a Gaussian prior $\tau = 0.055 \pm 0.009$. For the base-\LCDM\ model, using PR4 data, we obtain the same results as presented in~\citet{planck2016-l06}.  

We now describe the results obtained when fitting the tensor-to-scalar ratio $r$ in addition to the six \LCDM\ parameters ($\Omega_{\rm b}h^2$, $\Omega_{\rm c}h^2$, $\theta_\ast$, $A_{\rm s}$, $n_{\rm s}$, $\tau$). In \citet{planck2016-l10}, the constraint from $TT$ is reported as $r_{0.002} < 0.10$ (\CL{95}) using PR3 data. This is much lower than the expected 2$\,\sigma$ upper bound on $r$. Indeed, when we calculate $r_{\rm eff}$ as proposed in Sect.~\ref{sec:model}, we find that the maximum of the posterior is in the negative region by about 1.7$\,\sigma$.  
That the maximum happens to fall at negative values is the major reason for the apparently strong constraint on $r$.

\begin{figure}[htbp!]
	\begin{center}
	\includegraphics[draft=\draft,width=\columnwidth]{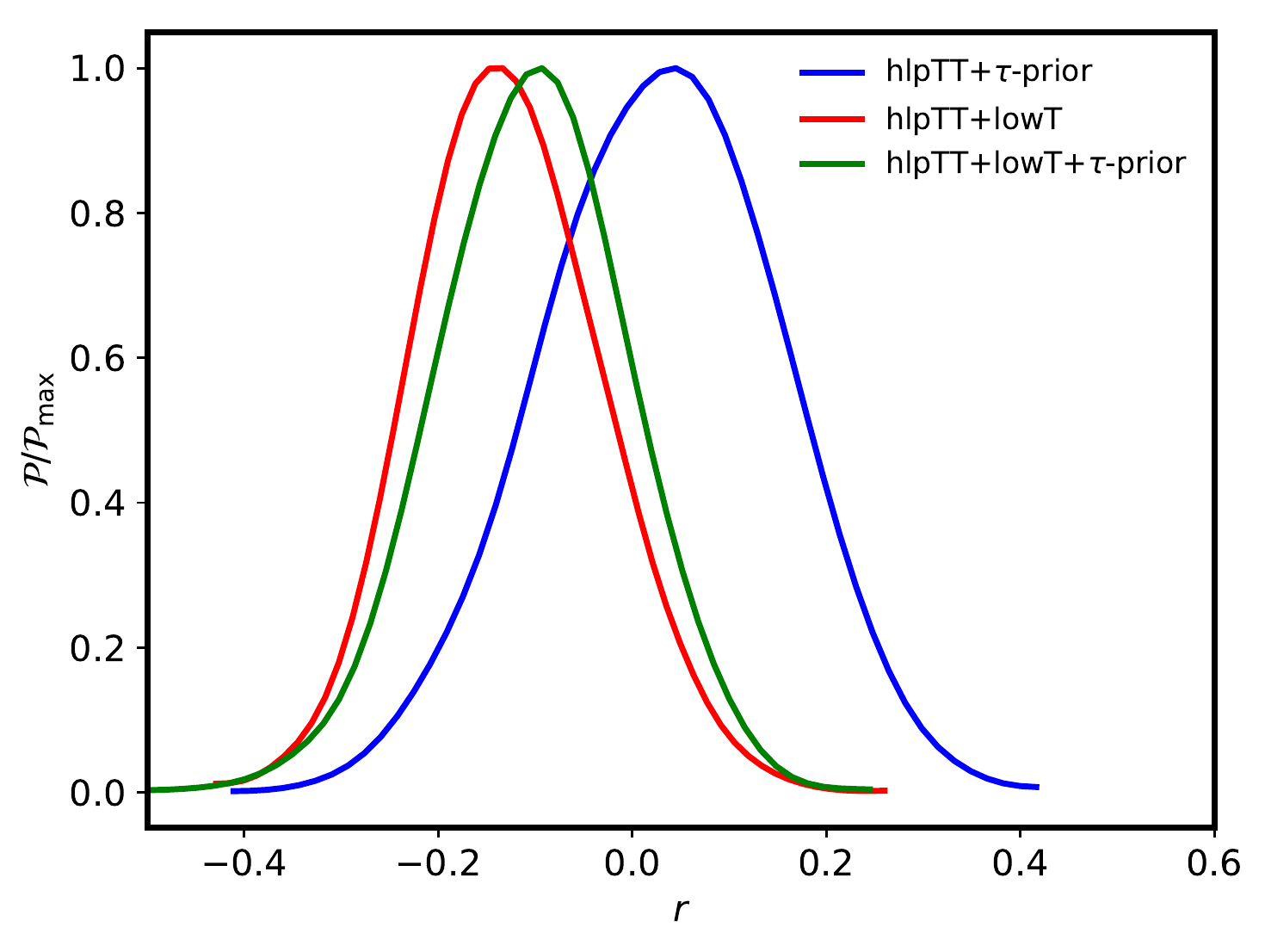}
	\caption{Constraints on the tensor-to-scalar ratio $r_{0.05}$ based on high-$\ell$ temperature data from \planck\ PR4 (\hlp TT), in combination with \lowT, and with a prior on $\tau$.}
	\label{fig:lik_r_TT_NPIPE}
	\end{center}
\end{figure}

With PR4 data, after marginalizing over the other cosmological parameters and the nuisance parameters, we find that the maximum of the posterior is negative by less than 1.2$\,\sigma$ when using \hillipop\ in temperature (\hlp TT) along with \lowT. As discussed in \citet{planck2016-l10}, this result is related to the \lowl\ deficit in the temperature power spectrum.  Indeed, removing \lowT\ from the likelihood moves the maximum of the posterior closer to zero, as illustrated in Fig.~\ref{fig:lik_r_TT_NPIPE}. The corresponding posterior maximum and 68\,\% confidence interval are
\begin{eqs}
	r_{0.05} &=& +0.031 \pm 0.120 \quad \text{(\hlp TT+$\tau$-prior)}, \\
	r_{0.05} &=& -0.131 \pm 0.093 \quad \text{(\hlp TT+\lowT)},\\
	r_{0.05} &=& -0.101 \pm 0.094 \quad \text{(\hlp TT+\lowT+$\tau$-prior)}.
\end{eqs}

Using the temperature power spectrum from PR4, we recover the same constraints on other parameters, in particular the scalar spectral tilt $n_{\rm s}$, as found using PR3 data (see Appendix~\ref{ann:PR3vsPR4}). With the full posterior distribution on $r$, we are able to accurately derive the maximum probability and the uncertainty $\sigma_r$. The width of the posterior is consistent with the PR3 results. Using only \highl\ data, with a prior on the reionization optical depth $\tau$, we find $\sigma_r = 0.12$ for $TT$ (consistent with the cosmic variance limit). Note that we find $\sigma_r = 0.43$ for $TE$, indicating that $TE$ is much less constraining for $r$ than $TT$. When adding information from low multipoles in temperature, $\sigma_r$ reduces to 0.094, but at the price of pushing the maximum distribution towards negative values. The posterior maximum is slightly shifted towards zero thanks to the small differences in \hillipop\ compared to the public \Planck\ likelihood (see Appendix~\ref{ann:hillipop}). The fact that the distribution peaks in the non-physical domain can be considered as a statistical fluctuation (with a significance between 1 and 2$\,\sigma$, depending on the data set used), which on its own is not a serious problem. However, the fact that this behaviour is strongly related to the deficit of power at \lowl\ in temperature is worth noting. 

After integrating the positive part of the $r$-posterior, the final upper limits from the \planck\ temperature power spectrum using PR4 are
\begin{eqs}
	r_{0.05} &<& 0.13 \quad  \text{(\CL{95}, \hlp TT+\lowT)},\\
	r_{0.05} &<& 0.12 \quad  \text{(\CL{95}, \hlp TT+\lowT+$\tau$-prior)}.
\end{eqs}

\section{Constraints from \textit{BB}}
\label{sec:bb}

To derive constraints on the tensor-to-scalar ratio from $BB$ using the PR4 maps, we sample the likelihood with a fixed \LCDM\ model based on the \Planck\ 2018 best fit, to which we add tensor fluctuations with a free amplitude parametrized by the tensor-to-scalar ratio $r$.
We use the \lollipop\ likelihood described in Sect.~\ref{sec:lik:lol}, restricted to $BB$ only (referred as `\lolB').  As discussed in Sect.~\ref{sec:lik:lol}, we construct the $C_\ell$ covariance matrix using the PR4 Monte Carlo simulations, which include CMB signal, foreground emission, realistic noise, and systematic effects.

Before giving the final constraints coming from the \planck\ $BB$ spectra, we should distinguish between the two different regimes, corresponding to large scales (the reionization bump) and intermediate scales (the recombination bump). Across the reionization bump, uncertainties are dominated by systematic residuals, as discussed in Sect.~\ref{sec:data}, while foreground residuals may bias the results. Across the recombination bump, uncertainties are dominated by statistical noise; however, systematic effects, as well as foreground residuals, can still bias constraints on $r$. In order to test the effects of potential foreground residuals, we calculate the posterior distributions of $r$ using various Galactic masks, as described in Sect.~\ref{sec:masks}. While large sky fractions ($f_{\rm sky} > 60\,\%$) show deviations from $r=0$, the posteriors for 40, 50, and 60\,\% of the sky are consistent with zero (Fig.~\ref{fig:lolR_galcuts}). As a robustness test, we also calculate the posterior distribution when changing the range of multipoles (Fig.~\ref{fig:lolR_lrange}) and find consistent results, with posteriors compatible with $r=0$. Multipoles above $\ell \simeq 150$ do not contribute to the result, since the noise in $BB$ is too high. For the rest of this paper, unless otherwise noted, we use a sky fraction of 50\%, and compute the likelihood over the range of multipoles from $\ell=2$ to $\ell=150$.

For the reionization and recombination bumps we find
\begin{eqnarray}
	r_{0.05} &=& -0.014_{-0.111}^{+0.108} \quad \text{(\lolB, reionization bump)}, \\
	r_{0.05} &=& \phantom{+}0.069_{-0.113}^{+0.114} \quad \text{(\lolB, recombination bump)}.
\end{eqnarray}
Both results are obtained over 50\,\% of the sky, with multipoles in the range $\ell = [2,35]$ for the former and $\ell=[50,150]$ for the latter. 
With these ranges of multipoles, and given the statistics of the PR4 maps, we can see that the reionization bump ($\sigma_r=0.110$) and the recombination bump ($\sigma_r=0.113$) contribute equally to the overall \planck\ sensitivity to the tensor-to-scalar ratio.

\begin{figure}[htbp!]
	\includegraphics[width=\columnwidth]{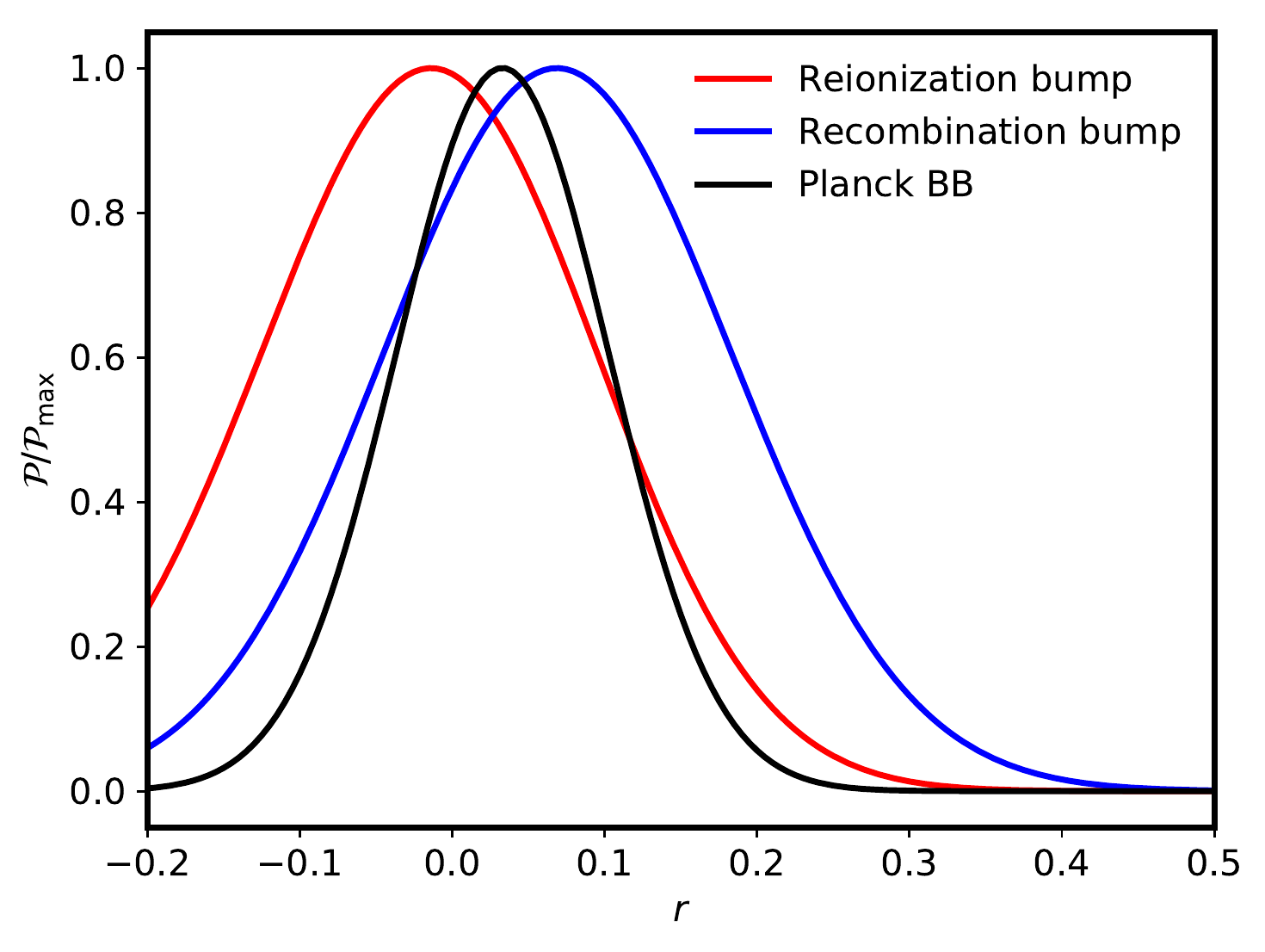}
	\caption{Posterior distribution of $r$ from PR4 data, using \lollipop\ and the $BB$ spectrum on 50\,\% of the sky (black).  Constraints from the reionization bump and the recombination bump are plotted in red and blue, respectively.  Constraints from \Planck\ BB with the full multipole range $\ell=[2,150]$ are in black.}
	\label{fig:lik_r_BB}
\end{figure}

We can combine the results from the two bumps in order to give the overall constraints on the tensor-to-scalar ratio from the \Planck\ $BB$ spectrum (Fig.~\ref{fig:lik_r_BB}).
The full constraint on $r$ from the PR4 $BB$ spectrum over 50\,\% of the sky, including correlations between all multipoles between $\ell=2$ and $\ell=150$,  is
\begin{eqnarray}
	r_{0.05} = 0.033 \pm 0.069 &&\quad \text{(\lolB)}.
\end{eqnarray}
This is fully compatible with no tensor signal, and we can derive an upper limit by integrating the posterior distribution out to 95\,\%, after applying the physical prior $r>0$, which yields
\begin{eqnarray}
	r_{0.05} < 0.158 &&\quad \text{(\CL{95}, \lolB)}.
\end{eqnarray}

This result can be compared with the BICEP2/Keck Array constraints \citep{Bicep2018limit} of
\begin{eqnarray}
	r_{0.05} < 0.072 &&\quad \text{(\CL{95}, BK15)}, 
\end{eqnarray}
with $\sigma_r=0.02$ compared to $\sigma_r = 0.069$ for the \Planck\ result presented in this analysis

\section{Additional constraints from polarization}
\label{sec:pol}

\begin{figure}[htbp!]
	\begin{center}
	\includegraphics[draft=\draft,width=\columnwidth]{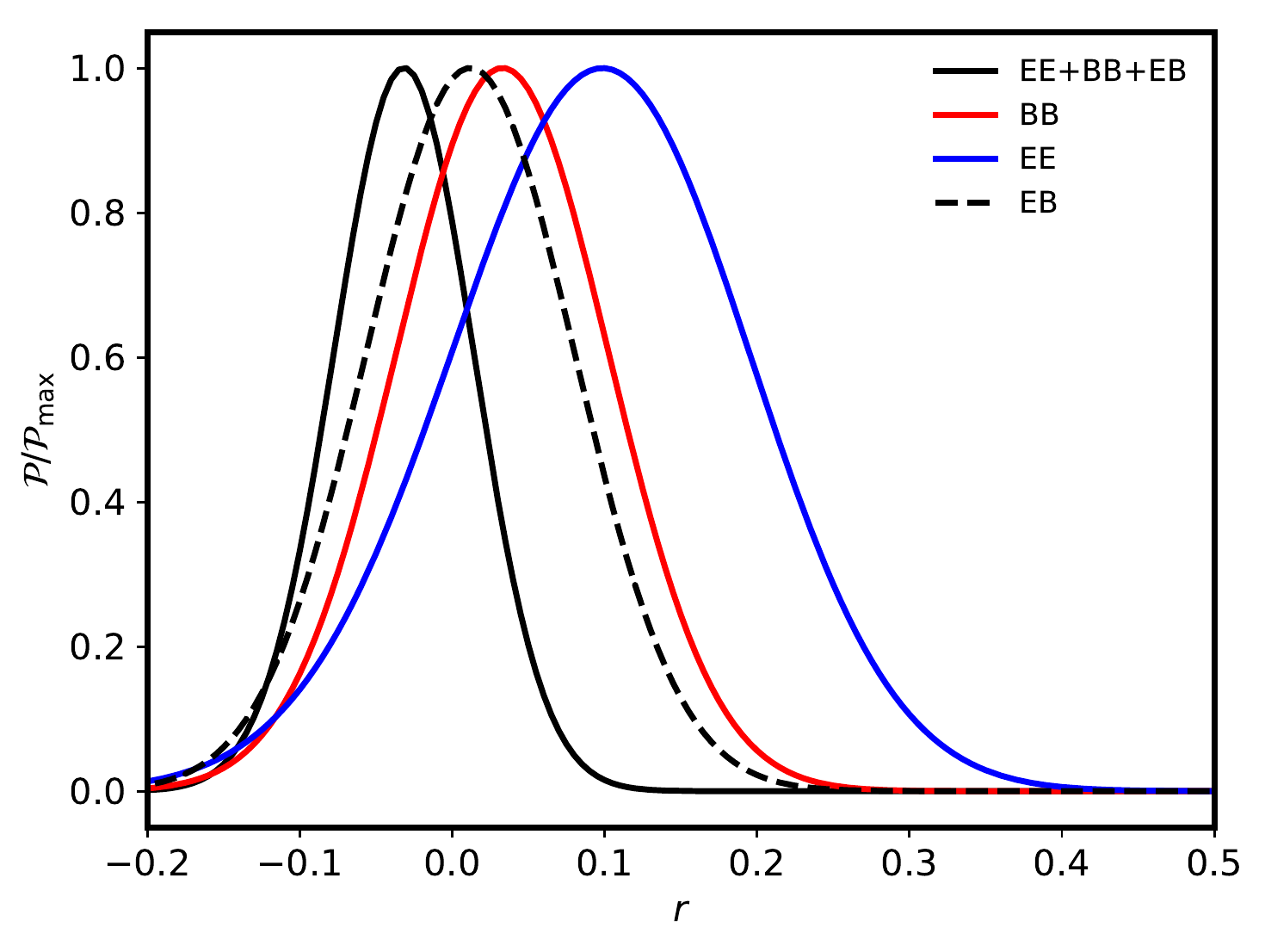}
	\caption{Posterior distributions for $r$ from \planck\ polarized \lowl\ data ($\ell < 150$) using \lollipop\ and the $EE$, $BB$, and $EE$+$BB$+$EB$ spectra.  The dashed black line is obtained from $EB$ data by fitting a $BB$ tensor model. The sky fraction used here is $f_{\rm sky} = 50\,\%$}
	\label{fig:lol_r_EB}
	\end{center}
\end{figure}

As shown in Fig.~\ref{fig:cl_tensor}, the $EE$ tensor spectrum is similar in amplitude to the $BB$ tensor spectrum, even though the scalar mode in $EE$ is stronger.  Given that noise dominates the tensor signal at all multipoles in both $EE$ and $BB$, we expect the likelihood for $EE$ to give useful constraints on $r$. We thus present the constraints from polarized \lowl\ data ($\ell < 150$) using different combinations of the \lollipop\ likelihood (specifically EE, BB, and EE+BB+EB) in Fig.~\ref{fig:lol_r_EB}.  We emphasize that EE+BB+EB is a likelihood of the correlated polarization fields $E$ and $B$ and not the combination of individual likelihoods (see Sect.~\ref{sec:lik:lol}).

The first thing to notice is that the posterior distribution for $EE$ peaks at $r = 0.098 \pm 0.097$, while the other modes give results compatible with zero within 1$\,\sigma$.
Given the lower sensitivity of \lolE\ to $r$ ($\sigma_r \simeq 0.10$) compared to that of \lolB\ ($\sigma_r \simeq 0.07$), this is mitigated when adding the information from other modes.
The posterior distributions for $r$ give
\begin{eqnarray}
	r_{0.05} = \phantom{+}0.033 \pm 0.069 &&\quad \text{(\lolB)},\\
	r_{0.05} = -0.031 \pm 0.046 &&\quad \text{(\lolEB)}.
\end{eqnarray}
As a consistency check, Fig.~\ref{fig:lol_r_EB} also shows the constraints when fitting the $BB$ tensor model on the $EB$ data power spectrum, which is compatible with zero ($r=-0.012 \pm 0.068$) as expected.

Using polarization data, \planck's sensitivity to the tensor-to-scalar ratio reaches $\sigma_r = 0.046$. Combining all \planck\ polarization modes ($EE$, $BB$, and $EB$) out to $\ell=150$ leads to the following upper limit:
\begin{eqnarray}
	r_{0.05} &<& 0.069 \quad \text{(\CL{95}, \lolEB)}.
\end{eqnarray}

\begin{figure}[htbp!]
	\begin{center}
	\includegraphics[draft=\draft,width=\columnwidth]{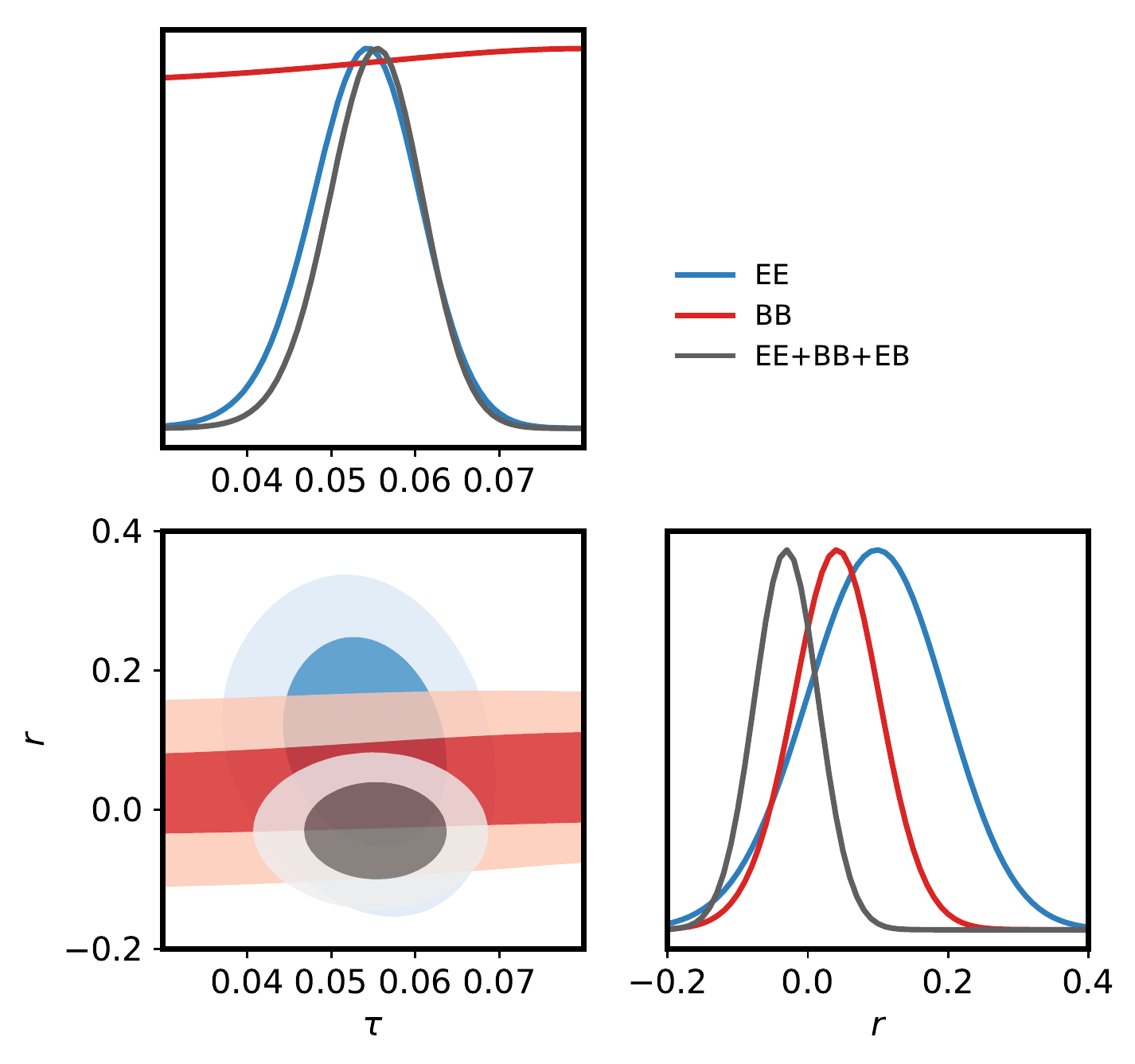}
	\caption{\lollipop\ posterior distribution in the $\tau$--$r$ plane using \lolE\ (blue), \lolB\ (red), and \lolEB\ (black). The sky fraction here is $f_{\rm sky} = 50\,\%$.}
	\label{fig:lol_tau-r}
	\end{center}
\end{figure}

Note that this constraint is almost independent of the other \LCDM\ parameters, and in particular the reionization optical depth $\tau$.
To demonstrate this, using the same data set (\lolB\ and \lolEB), we derive 2-dimensional constraints for $\tau$ and $r$ and plot them in Fig.~\ref{fig:lol_tau-r}.
The constraint is stable when sampling for $\tau$. Indeed, in this case, we obtain
\begin{eqnarray}
	r_{0.05} = \phantom{+}0.025 \pm 0.064 &&\quad \text{(\lolB)},\\
	r_{0.05} = -0.015 \pm 0.045 &&\quad \text{(\lolEB)},
\end{eqnarray}
and for the reionization optical depth
\begin{eqnarray}
	\tau = 0.0577 \pm 0.0056 &&\quad \text{(\lolEB)},
\end{eqnarray}
compatible with \lolE\ results, while \lolB\ shows no detection of $\tau$, since $BB$ is dominated by noise.

\section{Combined results}
\label{sec:combined}

Up to this point, the constraints on $r$ have been derived relative to a fixed fiducial \LCDM\ spectrum based on the \Planck\ 2018 results. Including the \planck\ temperature likelihoods (both \lowT\ and \hlp TT) in a combined analysis of the \planck\ CMB spectra allows us to properly propagate uncertainties from other cosmological parameters to $r$, as well as to self-consistently derive constraints in the $n_{\rm s}$--$r$ plane. 
In this section, we combine the \lowT\ and \hlp TT with the \lowl\ polarized likelihood \lolEB\ to sample the parameter space of the \LCDM+$r$ model.
The comparison of contours at 68\,\% and 95\,\% confidence levels between PR3 and PR4 data is presented in Fig.~\ref{fig:triangle_lcdm} of Appendix~\ref{ann:lcdm}.

We also include the BK15 constraints from \citet{Bicep2018limit}. When combining \Planck\ and BK15, we neglect the correlation between the two data sets and simply multiply the likelihood distributions. This is justified because the BK15 spectra are estimated on 1\,\% of the sky, while the \Planck\ analysis is derived from 50\,\% of the sky.

\begin{figure}[htbp!]
	\begin{center}
	\includegraphics[draft=\draft,width=\columnwidth]{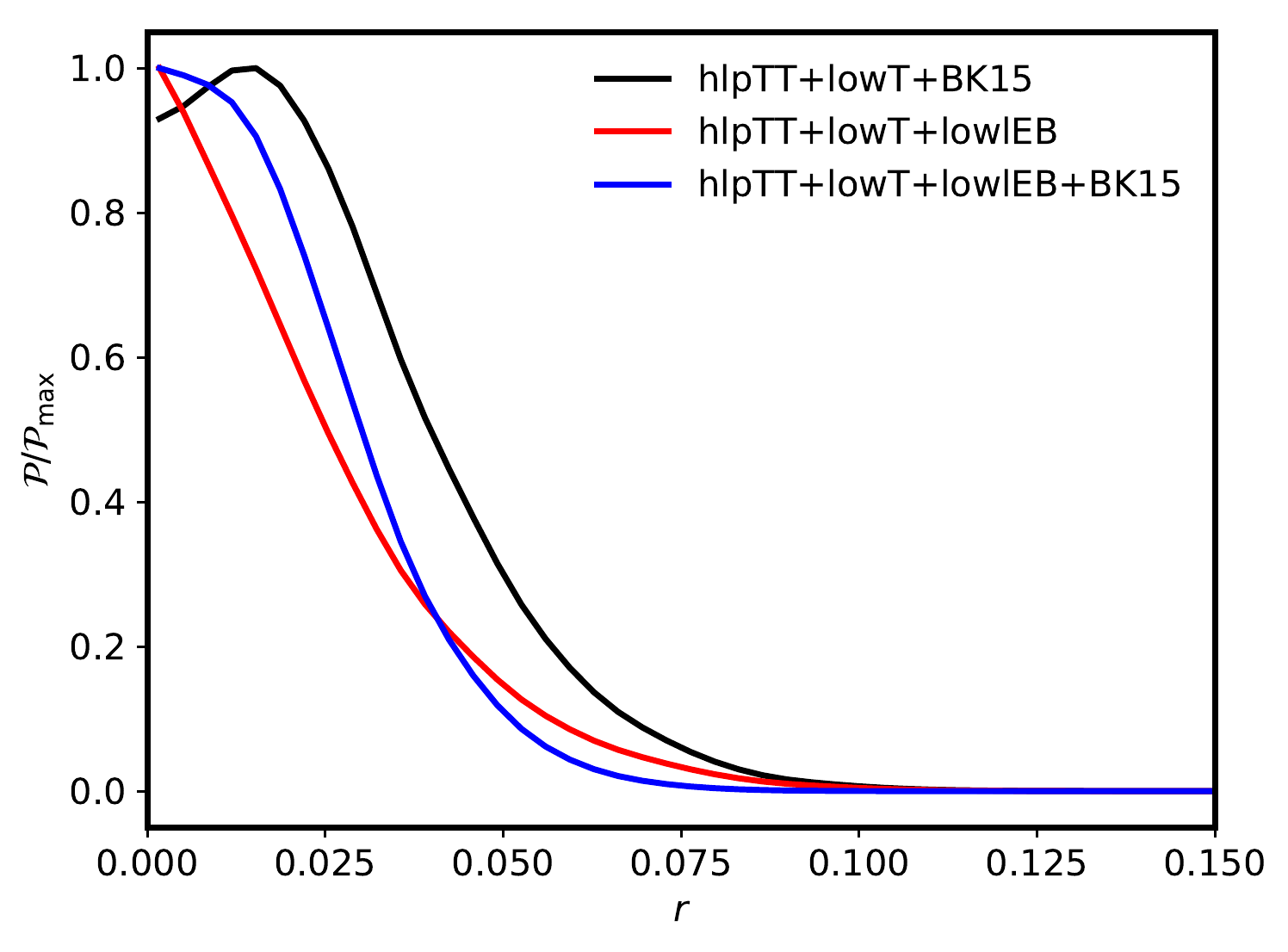}
	\caption{Posterior distributions for $r$ after marginalization over the nuisance parameters and the other \LCDM\ parameters, for the \planck\ temperature data (\hlp TT+\lowT) in combination with BK15 and the large-scale polarized \Planck\ likelihood (\lolEB).}
	\label{fig:likr_combined}
	\end{center}
\end{figure}

Figure~\ref{fig:likr_combined} gives posteriors on $r$ after marginalization over the nuisance and the other \LCDM\ cosmological parameters. We obtain the following \CL{95} upper limits:
\begin{eqs}
r_{0.05} &<& 0.060 \quad \text{(\CL{95}, \hlp TT+\lowT+BK15)};\\
r_{0.05} &<& 0.056 \quad \text{(\CL{95}, \hlp TT+\lowT+\lolEB)};\\
r_{0.05} &<& 0.044 \quad \text{(\CL{95}, \hlp TT+\lowT+\lolEB+BK15)}.
\end{eqs}

\begin{figure}[htbp!]
	\begin{center}
	\includegraphics[draft=\draft,width=\columnwidth]{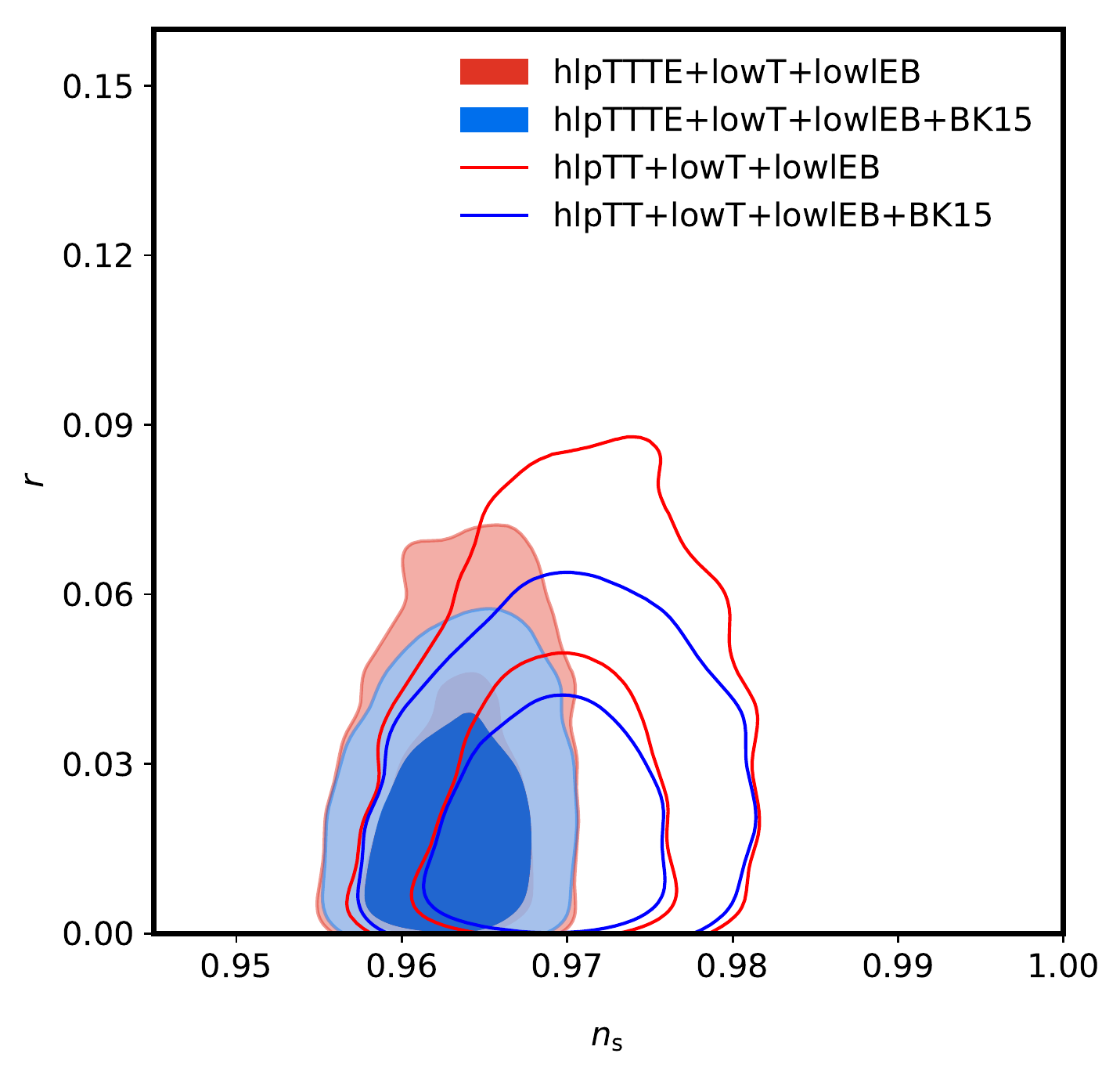}
	\caption{Marginalized joint 68\,\% and 95\,\% CL regions for $n_{\rm s}$ and $r_{0.05}$ from \Planck\ alone (\hlp+\lowT+\lolEB) and in combination with BK15. The solid lines correspond to \hlp TT+\lowT+\lolEB, while the filled regions include TE and correspond to \hlp TTTE+\lowT+\lolEB.}
	\label{fig:ns_r_inflation}
	\end{center}
\end{figure}

Figure~\ref{fig:ns_r_inflation} shows the constraints in the $r$--$n_{\rm s}$ plane for \planck\ data in combination with BK15. The constraints from the full combination of \Planck\ data are comparable to those from BK15.  The addition of the \highl\ $TE$ likelihood produces tighter constraints on the spectral index $n_{\rm s}$ \citep[as already reported in][]{planck2016-l06}.

There have been several other attempts to constrain the value of $r$, particularly through measurements of the $BB$ power spectrum.  As we have already stressed, there is a weak limit from the $TT$ spectrum and at the current sensitivity level for $r$, the constraints from $EE$ are about as powerful as those from $BB$; hence the tensor constraints in this paper are derived from a combination of $BB$ limits with those coming from $TT$ and $EE$. We show in Appendix~\ref{ann:BBplot} a comparison of our $BB$ limits with those of other experiments.

\section{Conclusions}

In this paper, we have derived constraints on the amplitude of tensor perturbations using \planck\ PR4 data. 
We investigated the intrinsic sensitivity of the $TT$ spectrum, which is cosmic-variance limited, and found $\sigma_r = 0.094$ using the full range of multipoles. We noted the impact of the \lowl\ anomaly, which pushes the maximum posterior distribution towards negative values of $r_{\rm eff}$ at roughly the 1$\,\sigma$ level.

For the first time, we analysed the \planck\ $BB$ spectrum for $r$ and obtained $\sigma_r = 0.069$, which is lower than in temperature.
The \planck\ $B$-mode spectrum, being dominated by noise, gives a constraint on $r$ that is fully compatible with zero from both low and intermediate multipoles, in other words from both the reionization and recombination peaks.  Multipoles above $\ell \simeq 150$ do not contribute to the result, since the noise in $BB$ is too high.

Using an appropriate likelihood in polarization, we showed that the \planck\ $EE$ spectrum is also sensitive to the amplitude of the tensor-to-scalar ratio $r$.
The combined constraints from \planck\ $EE$ and $BB$, including $EB$ correlations, lead to a sensitivity on $r$ of $\sigma_r = 0.046$, two times better than in temperature.
We also investigated the impact of foreground residuals using different Galactic cuts and by varying the range of multipoles used in the polarized likelihood.
Finally, by combining temperature and polarization constraints, we derived the posterior distribution on $r$ marginalized over the \LCDM\ cosmological parameters and nuisance parameters, including uncertainties from systematics (both instrumental and astrophysical). The result gives an upper limit of $r < 0.056$ at the 95\,\% confidence level using \planck\ data only.
In combination with the BICEP/Keck measurements from 2015, this constraint is further reduced to $r < 0.044$ (\CL{95}), the tightest limit on $r$ to date.

\begin{acknowledgements}
\Planck\ is a project of the European Space Agency (ESA) with instruments provided by two scientific consortia funded by ESA member states and led by Principal Investigators from France and Italy, telescope reflectors provided through a collaboration between ESA and a scientific consortium led and funded by Denmark, and additional contributions from NASA (USA).
Some of the results in this paper have been derived using the {\tt HEALPix} package.
This research used resources of the National Energy Research Scientific Computing Center (NERSC), a U.S. Department of Energy Office of Science User Facility operated under Contract No. DE-AC02-05CH11231.
We gratefully acknowledge support from the CNRS/IN2P3 Computing Center for providing computing and data-processing resources needed for this work.\end{acknowledgements}

\bibliographystyle{aat}
\bibliography{Planck_bib,planck_tensor}

\onecolumn
\appendix

\section{The \texttt{NPIPE} \textit{BB} transfer function}
\label{ann:bbtf}

The total signal simulations in the PR4 release have insufficient S/N to determine the CMB $BB$ transfer function \citep{planck2020-LVII} directly.  We therefore study the use of the well-measured CMB $EE$ transfer function in place of the unknown $BB$ function.  Such an approach is obviously approximate, but should be sufficient for an analysis providing an upper limit to the $BB$ amplitude, provided that the suppression of the $EE$ power in \NPIPE\ is \emph{at least} as strong as the $BB$ suppression.

\NPIPE\ filtering occurs during the destriping process, when time-domain templates are fitted against and subtracted from the TOD.  The filtering happens because the signal model in the calibration step does not include CMB polarization.  The processing approach mimics simple linear regression, except that it is performed in a subspace that does not include sky-synchronous degrees of freedom.

Here we set up a simplified test where we generate one survey's worth of single-detector, time-ordered data corresponding to pure CMB $E$ and pure CMB $B$ modes, and regress out all of the destriping templates, as visualized in appendix~E in \cite{planck2020-LVII}.  We then project both the unfiltered and filtered CMB signals into maps, and compute the `filtering function' for this test, which measures the amount of signal suppression between them as
\begin{equation}
  \label{eq:tf}
  f_\ell = C_\ell^{\,\mathrm{in}\times\mathrm{out}} / C_\ell^{\,\mathrm{in}\times\mathrm{in}}.
\end{equation}
For the purpose of this test, single-detector, single-survey data can be projected onto an intensity map, since the vast majority of the sky pixels are only observed in one orientation.  In Eq.~(\ref{eq:tf}), $C_\ell^{\,\mathrm{in}\times\mathrm{in}}$ is the pseudo-auto-spectrum of a map binned from the unfiltered signal outside a $\pm15^\circ$ cut in Galactic latitude, while $C_\ell^{\,\mathrm{in}\times\mathrm{out}}$ is the pseudo-cross-spectrum between the filtered and unfiltered maps.
This test produces a CMB $EE$ filtering function that resembles the magnitude and angular extent of the transfer function measured from the total signal simulations. This indicates that the missing elements of the data processing (full sky coverage, multiple surveys, multiple detectors, and projection operator) do not prevent us from drawing conclusions about the similarities between the $E$- and $B$-mode transfer functions.

\begin{figure*}[htbp!]
	\center
	\includegraphics[draft=\draft,width=0.92\textwidth]{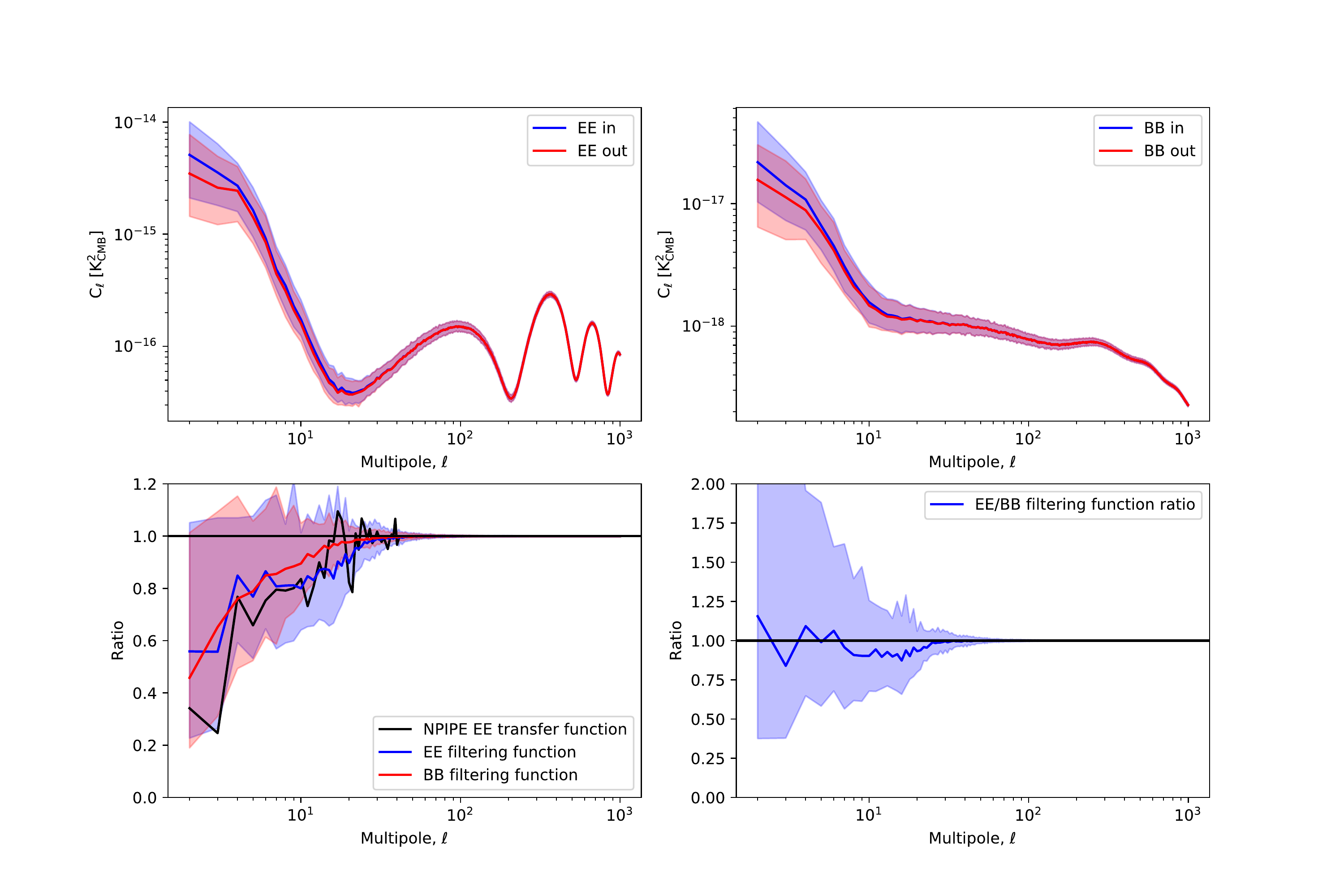}
	\caption{Linear regression test to compare coupling between CMB $E$ and $B$ modes and the \NPIPE\ destriping templates.
	\emph{Top:} Solid lines show the median bandpowers in individual multipole bins. The filled bands represent the 68\% confidence region around the median, as measured over 300 realizations. 
	\emph{Bottom left:} We compare measured $EE$ and $BB$ filtering functions derived from our simplified model with the $EE$ transfer function taken from PR4.
	\emph{Bottom right:} Ratio of the $EE$ and $BB$ filtering functions.}
  \label{fig:filtering_functions}
\end{figure*}

Results of $300$ Monte Carlo realizations are shown in Fig.~\ref{fig:filtering_functions}. They demonstrate good agreement between the $EE$ and $BB$ filtering functions and broad compatibility between the full simulation transfer function and our approximate test.  There is a statistically significant dip in the $EE/BB$ filtering function ratio in the range $\ell\simeq10$--20, suggesting that correcting the $BB$ spectra with the $EE$ transfer function might result in an overcorrection by a few percent, marginally inflating the upper limit. Given the approximate nature of this test and the modest size of the correction, we have not sought to tighten our upper limits by this difference.

\section{The \texttt{hillipop} likelihood}
\label{ann:hillipop}

\hillipop\ (High-$\ell$ Likelihood on Polarized Power-spectra) is one of the \highl\ likelihoods developed for analysis of the \Planck\ data. It was used as part of the 2013 \citep{planck2013-p08} and 2015 \citep{planck2014-a13} releases and also described in \citet{couchot2016} and \citet{couchot2017}.

\hillipop\ is very similar to the \Planck\ public likelihood (\plik). \hillipop\ is a Gaussian likelihood based on cross-spectra from the HFI 100-, 143-, and 217-GHz\ detset split-maps. The cross-spectra are estimated using a pseudo-$C_\ell$ algorithm with a mask adapted to each frequency to reduce the contamination from Galactic emission and point sources. The $C_\ell$ model includes foreground residuals on top of the CMB signal. These foreground residuals are both Galactic (dust emission) and extragalactic (CIB, tSZ, kSZ, SZ$\times$CMB, and point sources). \hillipop\ also introduced nuisance parameters to take into account map calibration uncertainties.

The most significant differences compared with \plik\ are that \hillipop\ uses:
\begin{itemize}
	\item detset-split maps instead of time splits (so that the cross-spectra do not need to account for a noise-correlation correction as in \plik);
	\item point-source masks that were obtained from a procedure that extracts compact Galactic structures;
	\item a Galactic dust $C_\ell$ model that (as a result of the mask difference) follows closely and is parametrized by the power law discussed in \citet{planck2014-XXX}, while \plik\ uses an ad-hoc effective function in $\ell$;
	\item foreground templates derived from \Planck\ measurements (\citealt{planck2014-XXX} for the dust emission, \citealt{planck2013-pip56} for the CIB, and \citealt{planck2014-a28} for the SZ) with a free amplitude for each emission mechanism, but spectral energy distributions fixed by \Planck\ measurements;
	\item a two-component model for the signal from unresolved point sources, which incorporates the contribution from extragalactic radio \citep{tucci11} and infrared dusty \citep{bethermin12} galaxies, as well as taking into account the variation of the flux cut across the sky and in `incompleteness' of the source catalogue at each frequency;
	\item all the 15 cross-spectra built from the 100-, 143-, and 217-GHz detset maps (while \plik\ keeps only five of them);
	\item all multipole values (while \plik\ bins the power spectra).
\end{itemize}

In the end, we have a total of six instrumental parameters (for map calibration) and nine astrophysical parameters (seven for $TT$, one for $TE$, and one for $EE$) in addition to the cosmological parameters (see Table~\ref{tab:hlp_nuisance}).

Using \hillipop\ on the PR3 data, we recover essentially identical constraints on the 6-parameter base-\LCDM\ model as the public \Planck\ likelihood. The results for the \LCDM+$r$ model with PR4 are discussed in Appendix~\ref{ann:PR3vsPR4}.

\begin{table}[htbp!]
\caption{Nuisance parameters for the \hillipop\ likelihood.}
\label{tab:hlp_nuisance}
\nointerlineskip
\vskip -3mm
\setbox\tablebox=\vbox{
    \newdimen\digitwidth
    \setbox0=\hbox{\rm 0}
    \digitwidth=\wd0
    \catcode`*=\active
    \def*{\kern\digitwidth}
    \newdimen\signwidth
    \setbox0=\hbox{+}
    \signwidth=\wd0
    \catcode`!=\active
    \def!{\kern\signwidth}
\halign{\hbox to 0.9in{#\leaderfil}\tabskip=2.0em&
    #\hfil\tabskip=2em&
    \hfil#\hfil\tabskip=0pt\cr
\noalign{\doubleline}
\omit\hfil Name\hfil&\omit\hfil Definition\hfil&\omit\hfil Prior (if any)\hfil\cr
\noalign{\vskip 3pt\hrule\vskip 4pt}
\multicolumn{3}{c}{Instrumental}\cr
\noalign{\vskip 3pt\hrule\vskip 4pt}
$c_0$ & map calibration (100-A) &  $0.000 \pm 0.002$\cr
$c_1$ & map calibration (100-B) &  $0.000 \pm 0.002$\cr
$c_2$ & map calibration (143-A) &  fixed\cr
$c_3$ & map calibration (143-B) &  $0.000 \pm 0.002$\cr
$c_4$ & map calibration (217-A) &  $0.000 \pm 0.002$\cr
$c_5$ & map calibration (217-B) &  $0.000 \pm 0.002$\cr
$A_{\rm pl}$ & absolute calibration & $1 \pm 0.0025$\cr
\noalign{\vskip 3pt\hrule\vskip 4pt}
\multicolumn{3}{c}{Foreground modelling}\cr
\noalign{\vskip 3pt\hrule\vskip 4pt}
$A_{\rm PS}^{\rm radio}$        & scaling parameter for radio sources in TT                      &\cr
$A_{\rm PS}^{\rm IR}$   & scaling parameter for IR sources in TT                                &\cr
$A_{\rm SZ}$                    & scaling parameter for the tSZ in TT                           &\cr
$A_{\rm CIB}$                   & scaling parameter for the CIB in TT                           & $1.00 \pm 0.20$\cr
$A_{\rm dust}^{\rm TT}$         & scaling parameter for the dust in TT                          & $1.00 \pm 0.20$\cr
$A_{\rm dust}^{\rm EE}$         & scaling parameter for the dust in EE                          & $1.00 \pm 0.20$\cr
$A_{\rm dust}^{\rm TE}$         & scaling parameter for the dust in TE                          & $1.00 \pm 0.20$\cr
$A_{\rm kSZ}$                   & scaling parameter for the kSZ effect in TT                      &\cr
$A_{{\rm SZ}\times{\rm CIB}}$         & scaling parameter for SZ$\times$CIB in TT     &\cr
\noalign{\vskip 4pt\hrule\vskip 3pt}}}
\endPlancktablewide
\end{table}

\section{Large-scale polarized angular power spectra}
\label{ann:pse}

We estimate CMB large-scale polarized power spectra by cross-correlating the two independent detset splits, A and B, after \commander\ component separation.
As detailed in Sect.~\ref{sec:lik:lol}, we use two different power-spectrum estimators for the low ($2 \leq \ell \leq 35$) and the intermediate ($35<\ell<300$) multipole ranges. For the lower multipole range we use a quasi-QML estimator \citep{vanneste18}, while for the higher multipoles we use a classic pseudo-$C_\ell$ approach \citep[a generalization to polarization of the method presented in][]{tristram2005} with a binning of $\Delta\ell = 10$.

\begin{figure*}[htbp!]
	\centering
	\includegraphics[height=250pt]{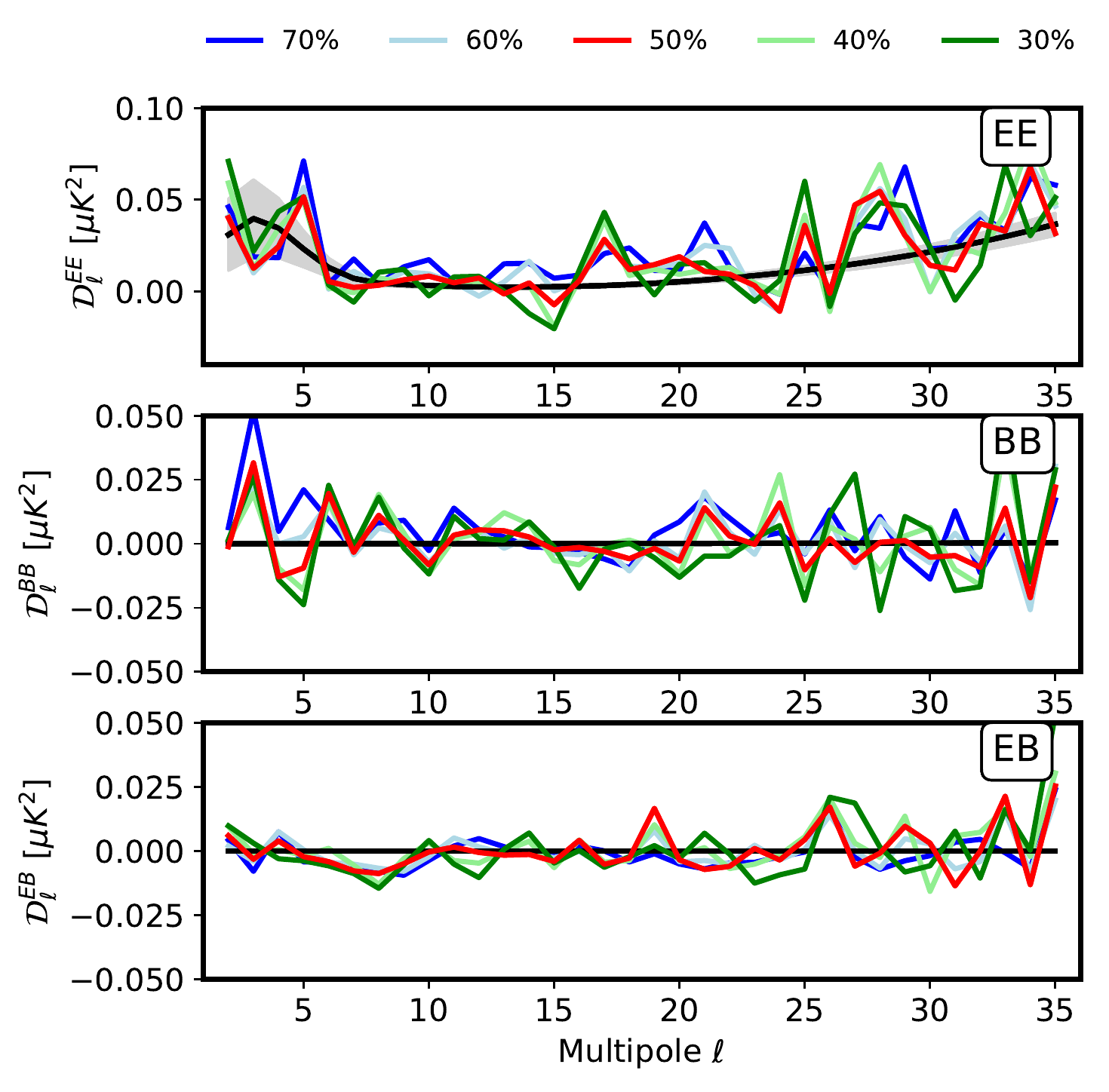}
	\includegraphics[height=250pt]{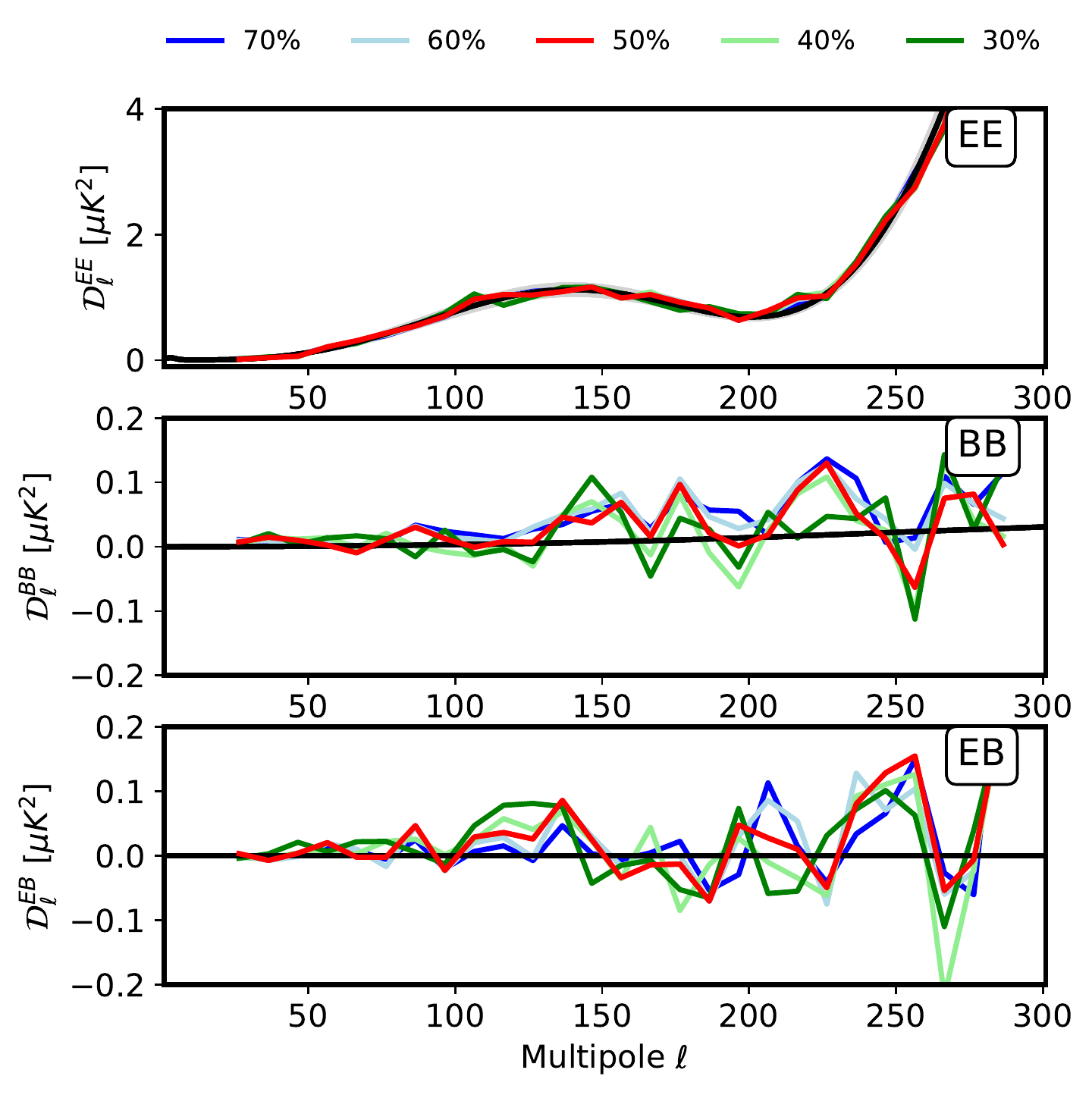}
	\caption{$EE$, $BB$, and $EB$ power spectra computed from the PR4 maps for sky fractions from 30 to 70\,\%.  The black lines represent the \LCDM\ model and the grey bands show its associated full-sky cosmic variance.}
	\label{fig:cl_galcut}
\end{figure*}

Figure~\ref{fig:cl_galcut} shows the reconstructed $EE$, $BB$, and $EB$ power spectra for various sky fractions, from 30 to 70\,\%. The 6-parameter \LCDM\ model based on the best fit to the \planck\ 2018 data is plotted in black, together with the cosmic variance (computed for the full sky) in grey.
The spectra show a remarkable consistency with the model, and are largely insensitive to sky fraction.  The $BB$ and $EB$ spectra are dominated by noise. At low multipoles, the $BB$ spectrum computed on the largest sky fraction (70\,\%) exhibits an excess at very low multipoles ($\ell \le 5$), attributable to Galactic residuals. For both $BB$ and $EB$, a reduction of the sky fraction corresponds to a larger dispersion.

\vspace{-0.3cm}

\begin{figure*}[htbp!]
	\centering
	\includegraphics[height=250pt]{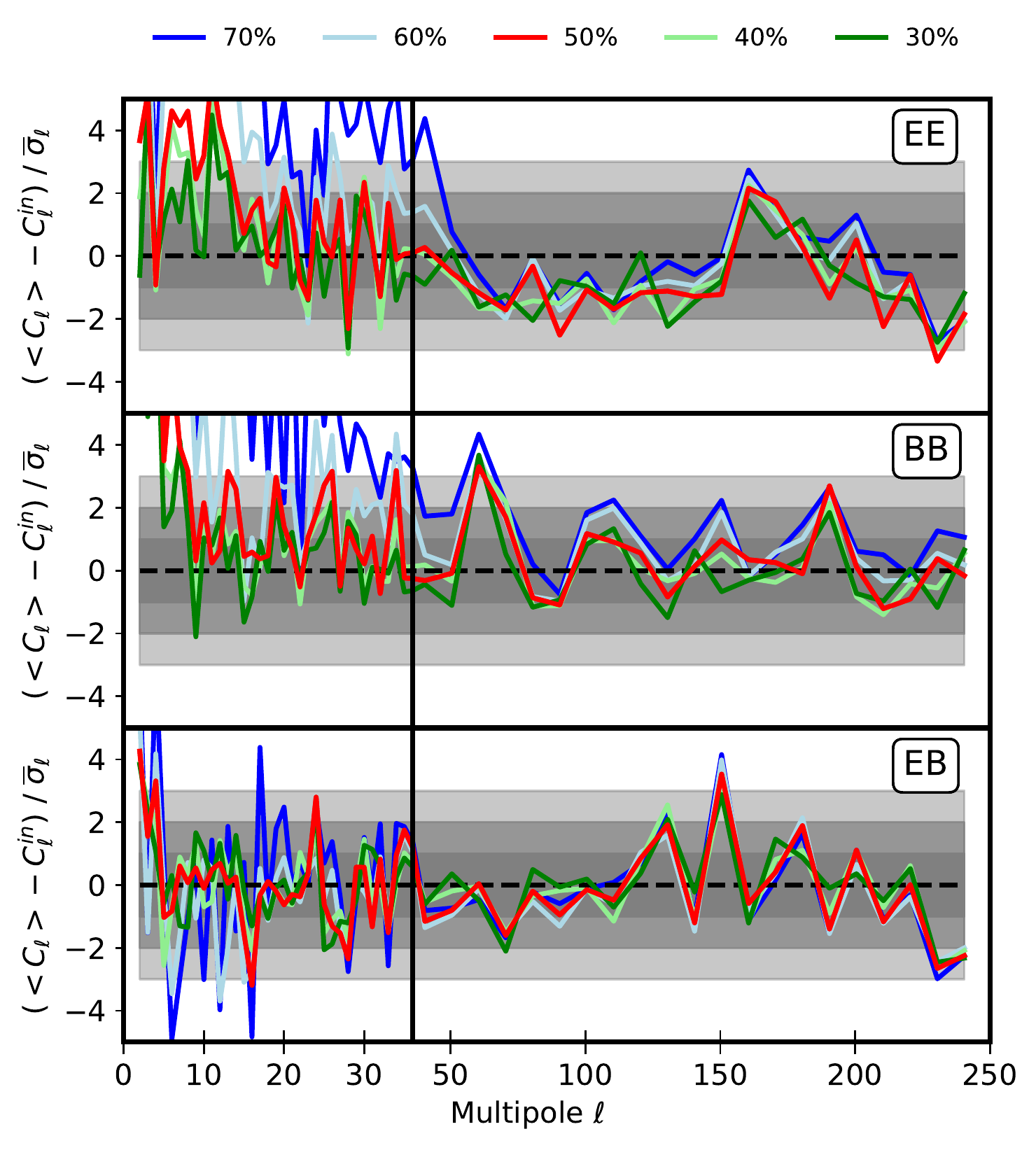}
	\includegraphics[height=250pt]{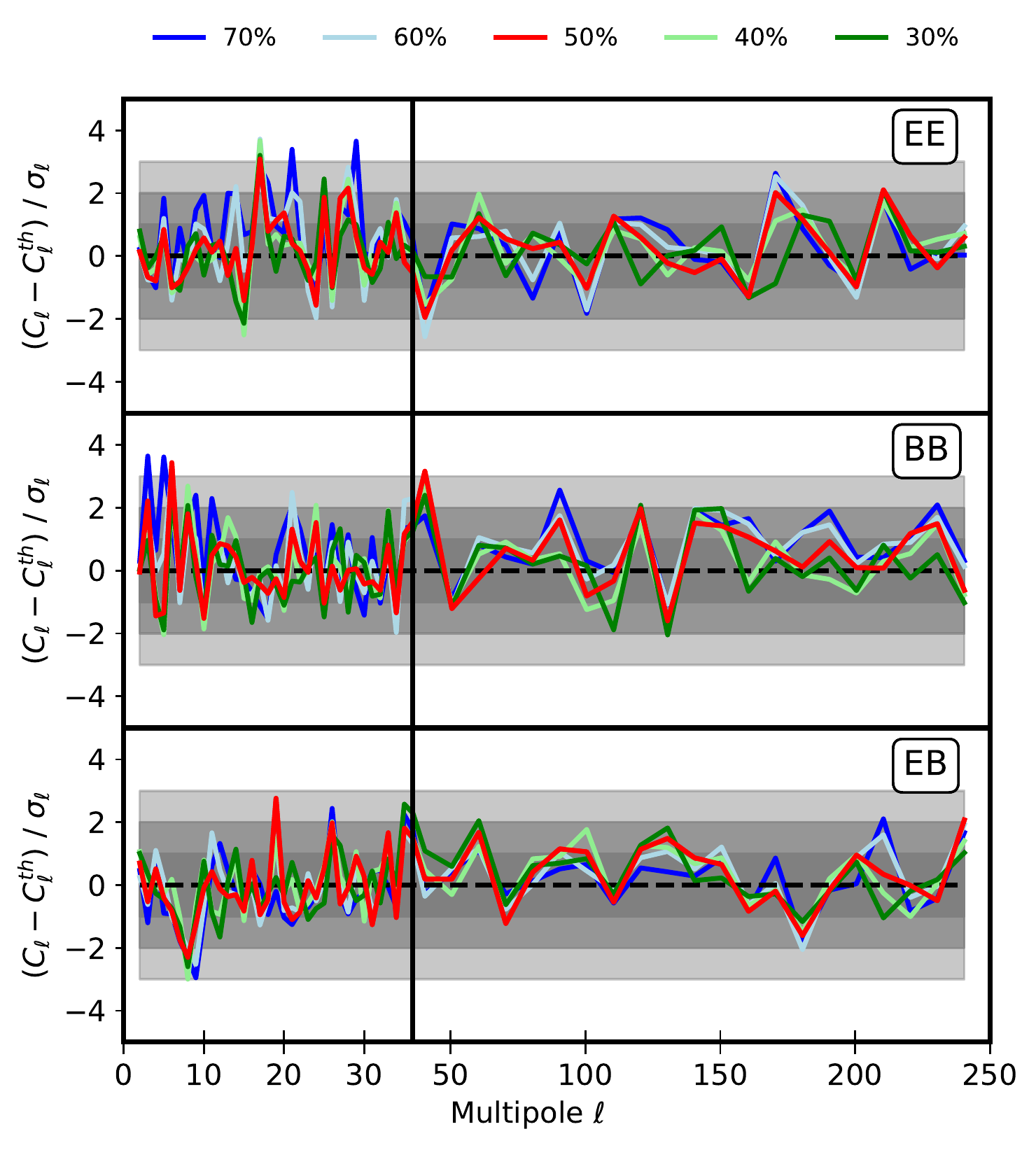}
	\caption{Residuals of the $EE$, $BB$, and $EB$ power spectra. {\it Left:} Mean value of the Monte Carlo simulations compared to the error on the mean, $\overline{\sigma}_\ell = \sigma_\ell\,n_{\rm sim}^{-1/2}$. \, {\it Right:} PR4 data compared to the standard deviation $\sigma_\ell$.  Grey bands show the 1, 2, and $3\,\sigma$ levels.}
	\label{fig:cl_residuals}
\end{figure*}

Figure~\ref{fig:cl_residuals} shows the residuals for the $EE$, $BB$, and $EB$ power spectra compared to the best-fit base-\LCDM\ model from the \Planck\ 2018 results. The plot on the left shows the residuals from the Monte Carlo simulations computed as the average spectra over the simulations divided by the uncertainty in the mean, $\overline{\sigma}_\ell = \sigma_\ell\,n_{\rm sim}^{-1/2}$. The average here is computed independently for each multipole.
The plot on the right shows the residuals of the PR4 data compared to the \LCDM\ model divided by the spectrum uncertainties (i.e., the dispersion over the simulations).

\vspace{-0.3cm}

\section{Cross-spectrum correlation matrix}
\label{ann:corrmat}

Uncertainties are propagated to the likelihood function through the $C_\ell$ covariance matrices (see Sect.~\ref{sec:lik:lol}).  We use Monte Carlo simulations to estimate the covariance over the multipoles considered in this analysis ($2 \leq \ell \leq 150$). 
Four hundred simulations are available, allowing us to invert the estimated covariance matrices of rank 45 for $BB$ and 135 for the full $EE$+$BB$+$EB$. While the uncertainties would be better determined with a larger sample of simulations, we have seen no impact on the results when estimating the covariance matrix on a reduced number of simulations, and we conclude that our sample of 400 is enough to propagate the uncertainties of the $C_\ell$s to $r$, as well as to $\tau$.

\begin{figure*}[htbp!]
	\centering
	\includegraphics[width=.495\columnwidth]{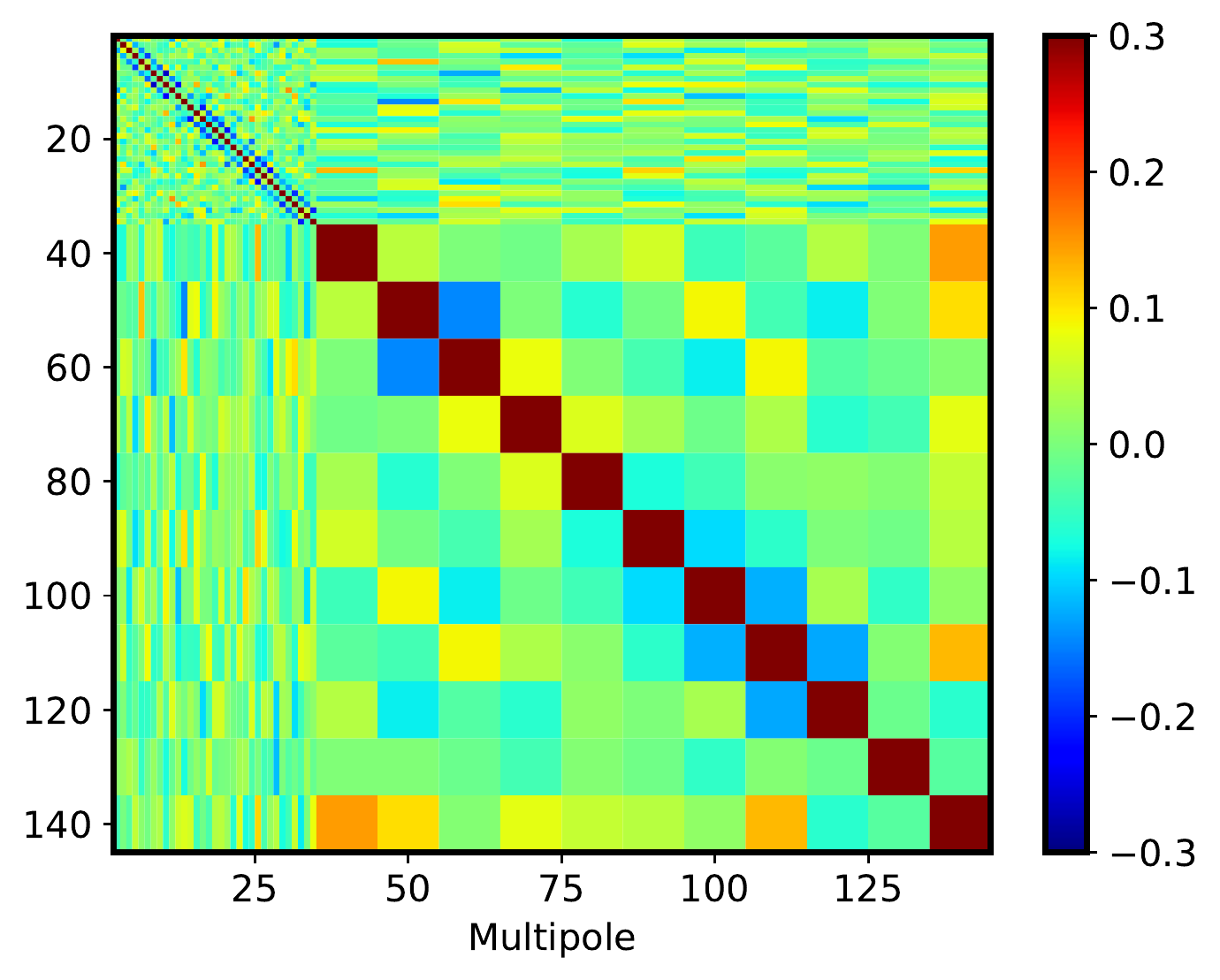}
	\includegraphics[width=.495\columnwidth]{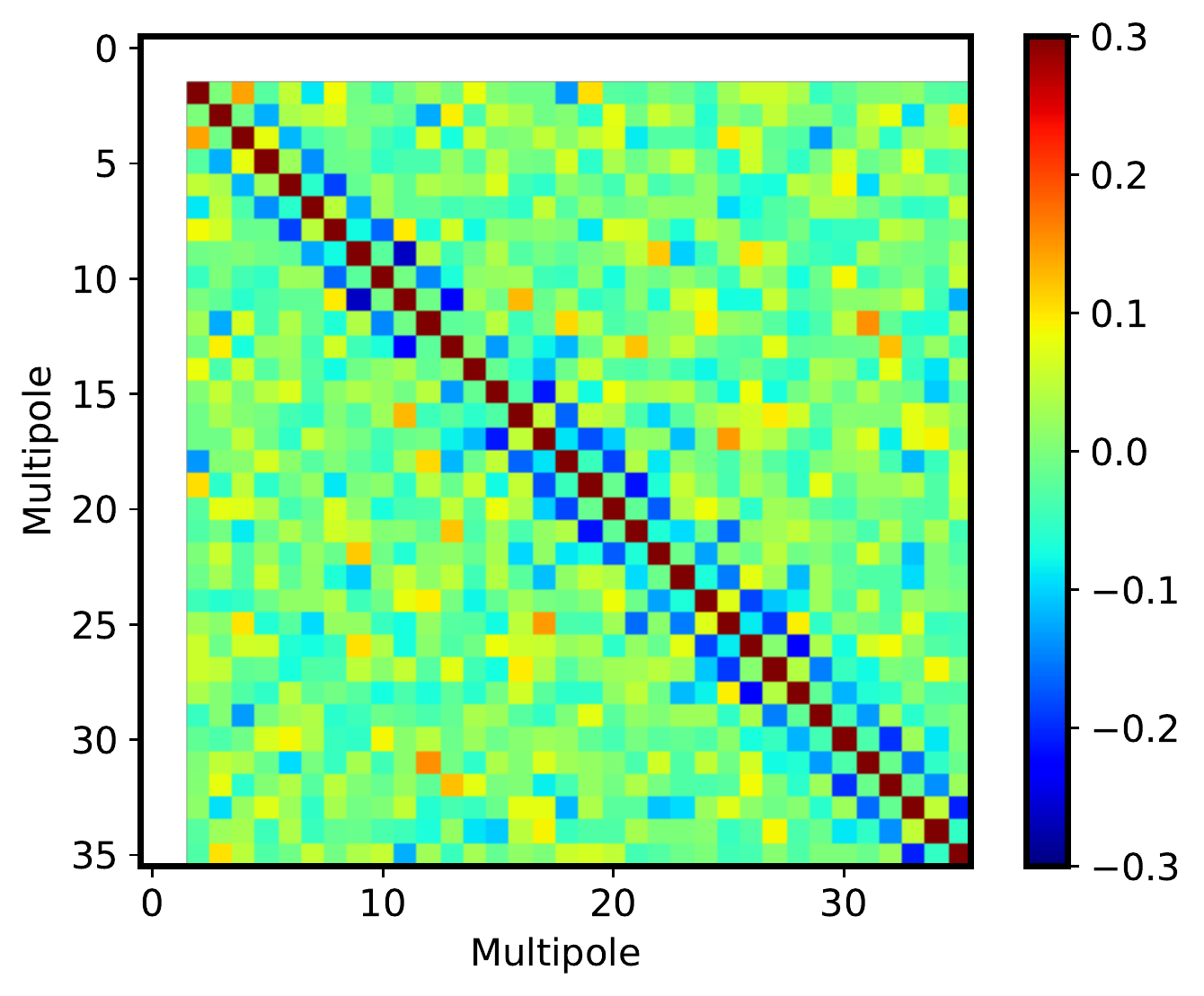}
	\caption{{\it Left:} Correlation matrix for $C_\ell^{BB}$. \, {\it Right:} Enlargement of the upper-left corner.  Covariances are estimated from 400 end-to-end simulations, including signal, noise, systematics, and foreground residuals.}
	\label{fig:corrmatBB}
\end{figure*}

\begin{figure*}[htbp!]
	\centering
	\includegraphics[width=0.65\columnwidth]{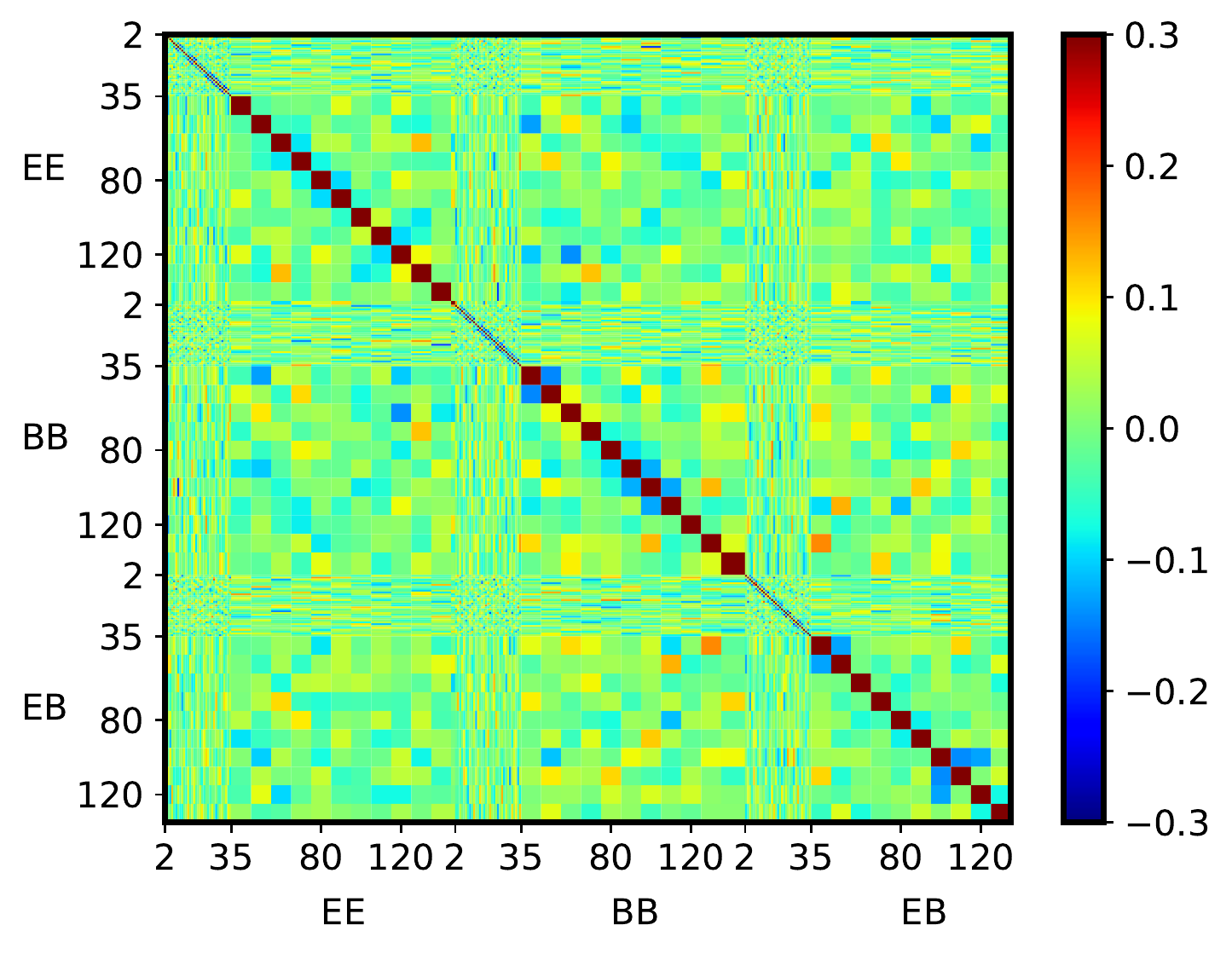}
	\caption{Correlation matrix for the $C_\ell^{EE}$, $C_\ell^{BB}$, and $C_\ell^{EB}$. Covariances are estimated from 400 end-to-end simulations, including signal, noise, systematics, and foreground residuals.}
	\label{fig:corrmat}
\end{figure*}

Figure~\ref{fig:corrmatBB} shows the correlations of the $C_\ell^{BB}$ computed from the covariance matrix.  Correlations from bin to bin are below 15\,\%, except for the next-to-neighbour bins at $\ell < 40$.
Figure~\ref{fig:corrmat} shows the correlations computed from the full covariance matrix, including $EE$, $BB$, and $EB$ spectra for all multipole bins from 2 to 150.

\section{$\mathbf{\Lambda}$CDM+\textit{r} parameters for PR3 and PR4}
\label{ann:PR3vsPR4}
\begin{figure*}[htbp!]
	\center
	\includegraphics[width=0.99\textwidth]{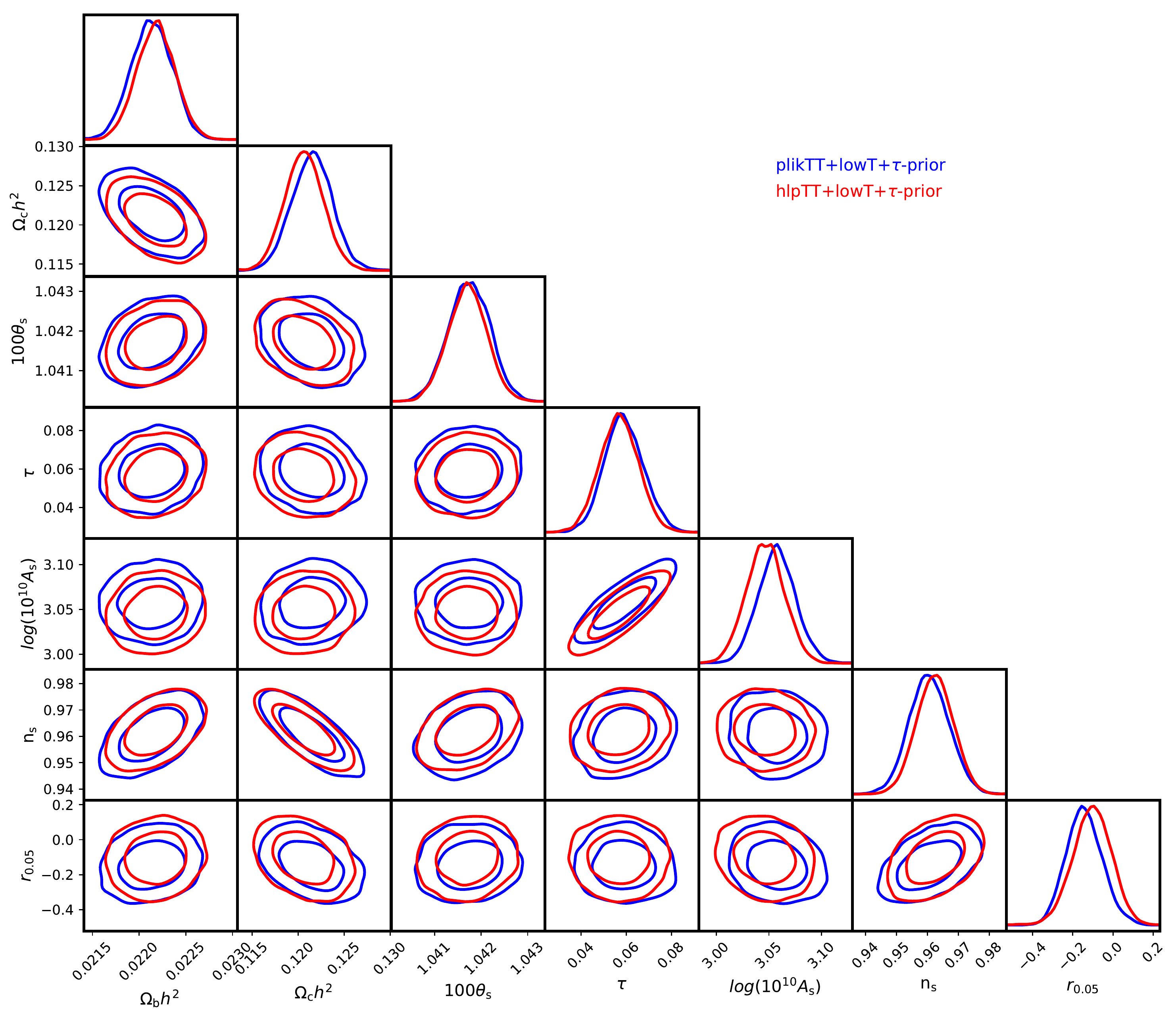}
	\caption{Constraint contours (at 68 and 95\,\% confidence) on parameters of a \LCDM+$r$ model using PR4 and \hillipop\ (red), compared to those obtained with PR3 and \plik\ (blue).  Both sets of results use the \commander\ likelihood for temperature multipoles $\ell \leq 30$, and a Gaussian prior to constrain $\tau$.}
	\label{fig:triangle_lcdm_TT}
\end{figure*}

In this paper, we use a new data set, PR4, and an alternative likelihood, \hillipop. 
Figure~\ref{fig:triangle_lcdm_TT} shows that we obtain essentially the same parameter values for the \LCDM+$r$ model based on the {\it temperature\/} power spectrum as are obtained from PR3 and \plik\ (available on the Planck Legacy Archive).  The differences between PR3 and PR4 are primarily seen in polarization.  

\section{Robustness tests}
\label{ann:robustness}

\begin{figure*}[htbp!]
	\center
	\subfigure[\lolB]{
	\includegraphics[draft=\draft,width=0.48\textwidth]{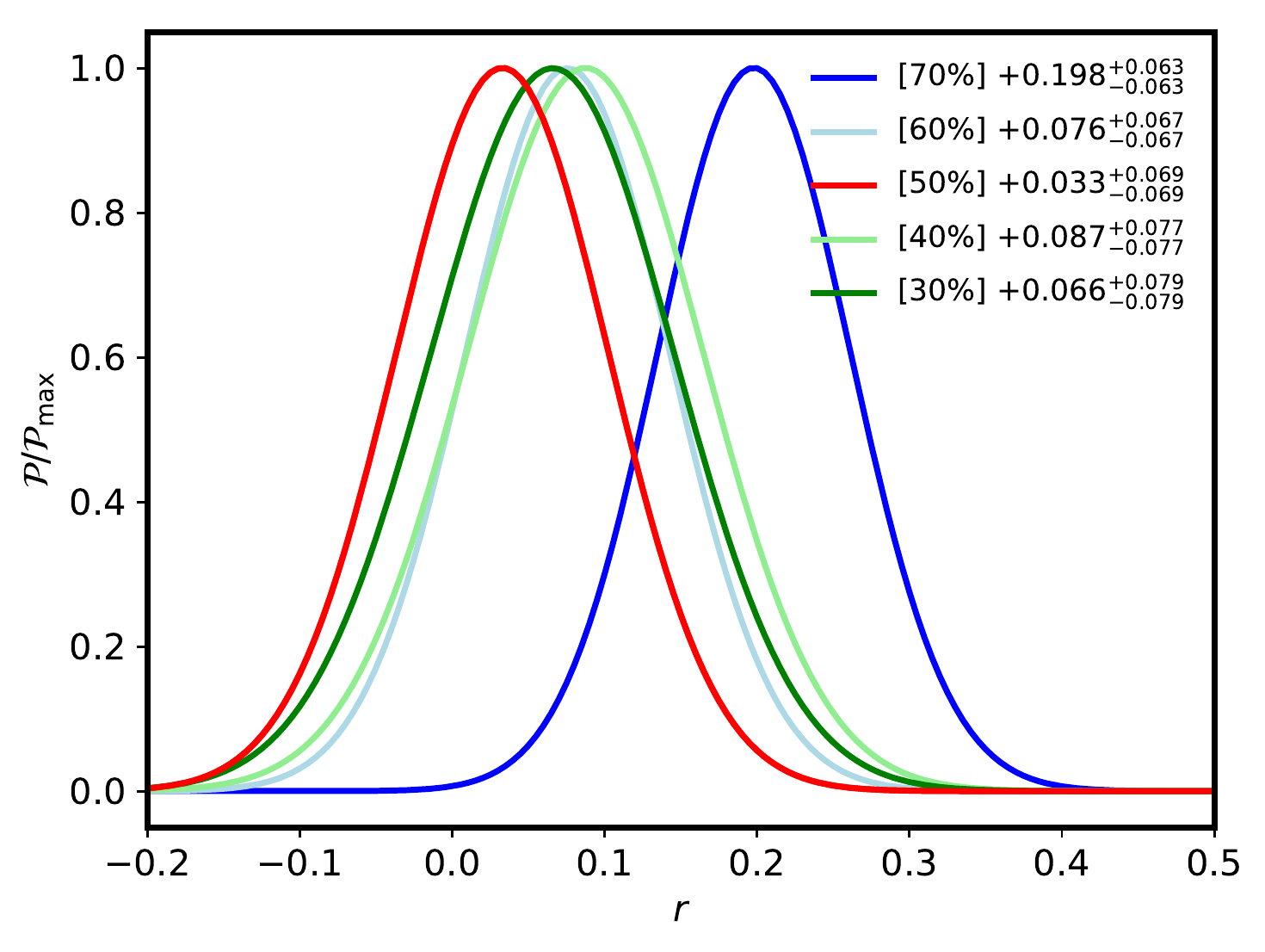}
	}
	\subfigure[\lolEB]{
	\includegraphics[draft=\draft,width=0.48\textwidth]{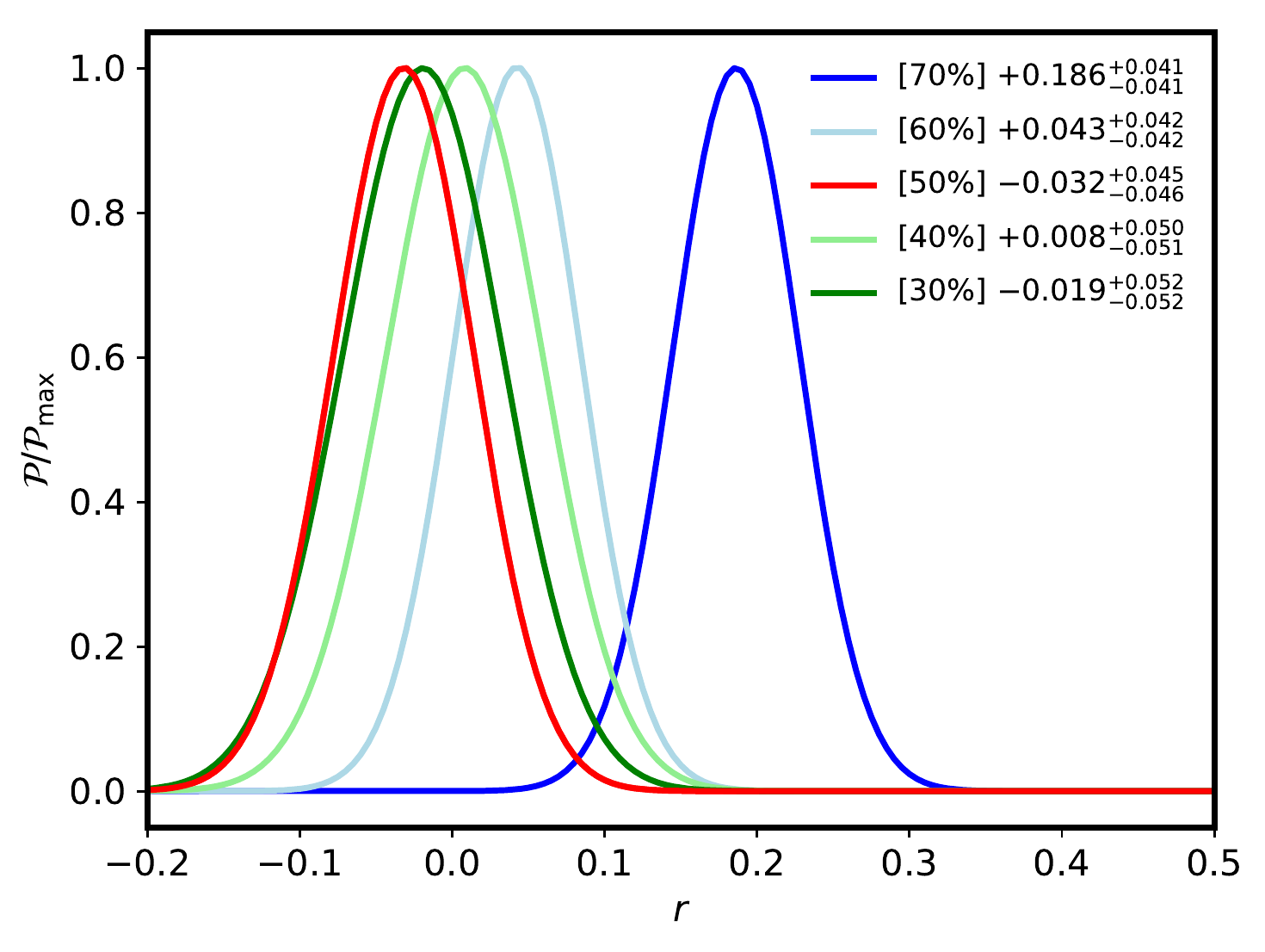}
	}
	\caption{Posterior distributions for r using \lollipop\ and $BB$ (panel a) and $EE+BB+EB$ (panel b) for sky fractions from 30\,\% to 70\,\%.  The multipole range is the same in all cases, $\ell = [2,145]$.  The best combination of sensitivity and freedom from foreground residuals is achieved at 50\,\%.}
	\label{fig:lolR_galcuts}
\end{figure*}

\begin{figure*}[htbp!]
	\center
	\subfigure[$\ell_{\rm min}$]{
	\includegraphics[draft=\draft,width=0.48\textwidth]{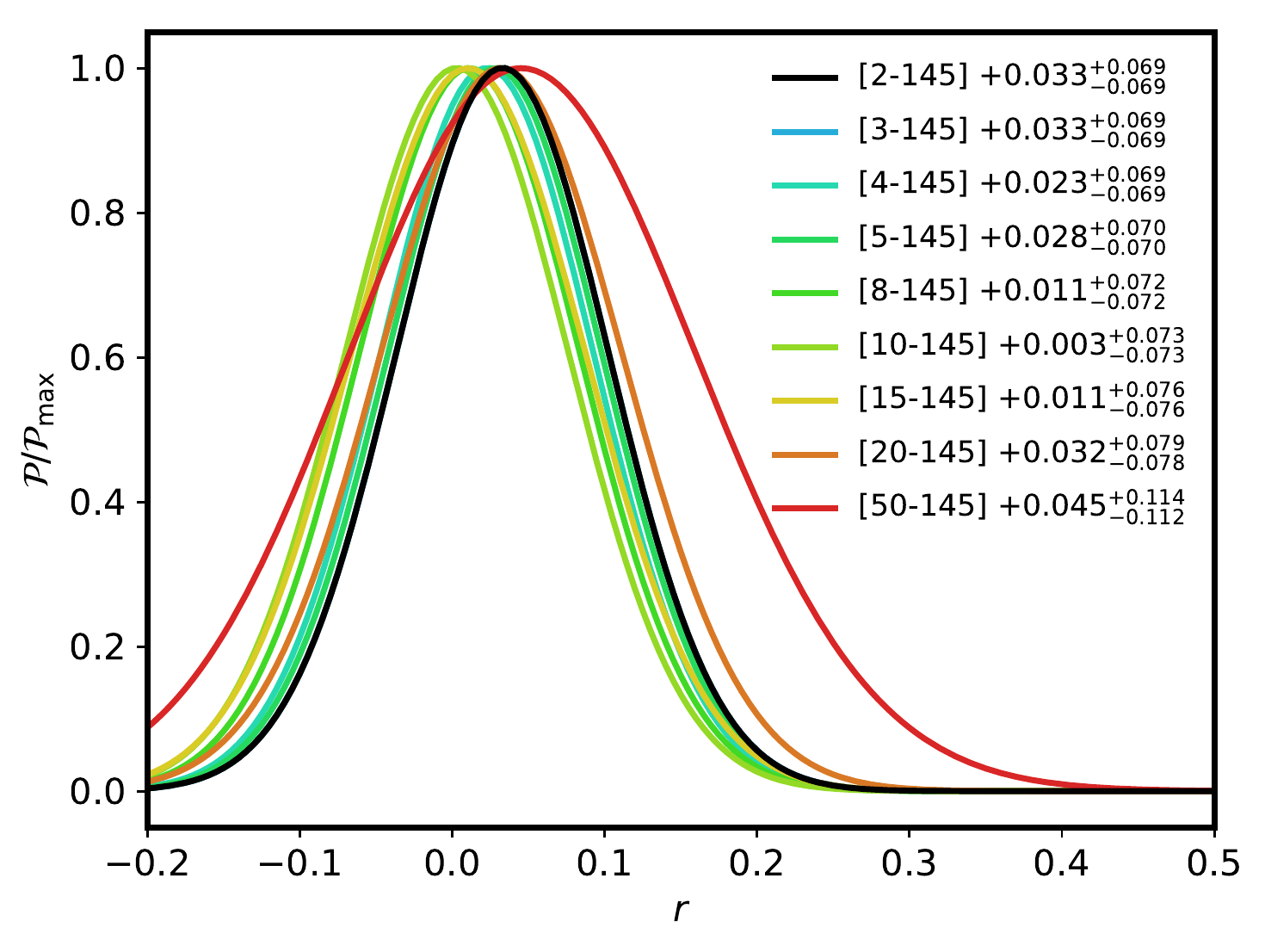}
	}
	\subfigure[$\ell_{\rm max}$]{
	\includegraphics[draft=\draft,width=0.48\textwidth]{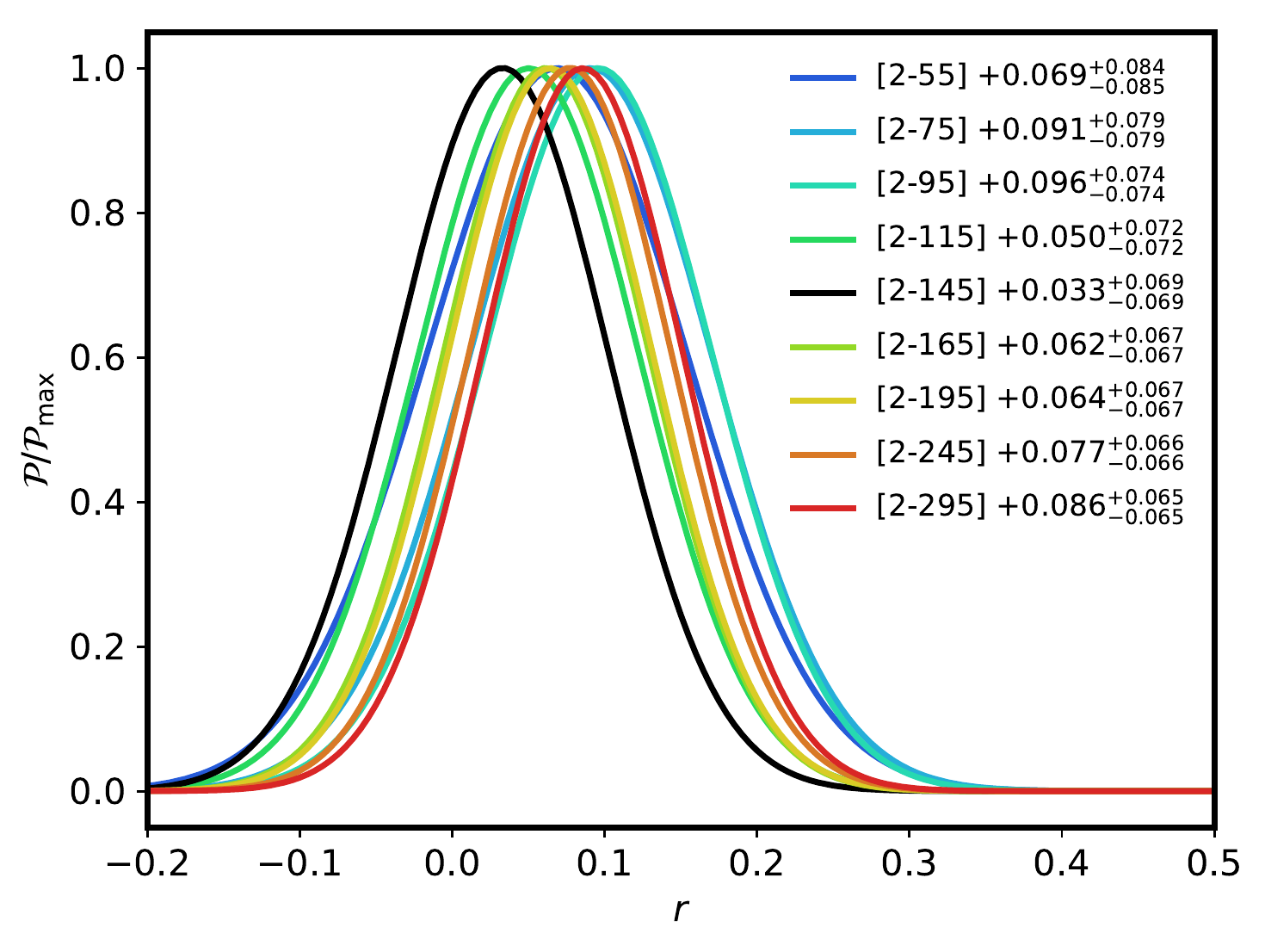}
	}
	\caption{Posterior distributions for $r$ using \lollipop\ and the $BB$ power spectrum for various multipole ranges. The sky fraction is fixed at 50\,\%.}
	\label{fig:lolR_lrange}
\end{figure*}

As a test of robustness, we computed the posterior distributions for $r_{\rm eff}$ considering spectra estimated on different sky fractions from 30 to 70\,\%  (Sect.~\ref{sec:masks}).  Figure~\ref{fig:lolR_galcuts} shows results for the BB (left) and EE+BB+EB (right) likelihoods.
A sky fraction of 50\,\% provides the best combination of sensitivity (which increases with the sky fraction) and freedom from foreground residuals (which decreases with sky fraction).
Figure~\ref{fig:lolR_lrange} shows that the BB posterior is robust with respect to the choice of $\ell_{\rm min}$  up to $\ell_{\rm min} \approx 20$, and that the width of the posterior is stable for $\ell_{\rm max} \gsim 150$, a consequence of the increase of signal-to-noise ratio with multipole.

\
\section{Triangle plot for $\mathbf{\Lambda}$CDM+\textit{r} parameters}
\label{ann:lcdm}

\begin{figure*}[htbp!]
	\includegraphics[draft=\draft,width=\textwidth]{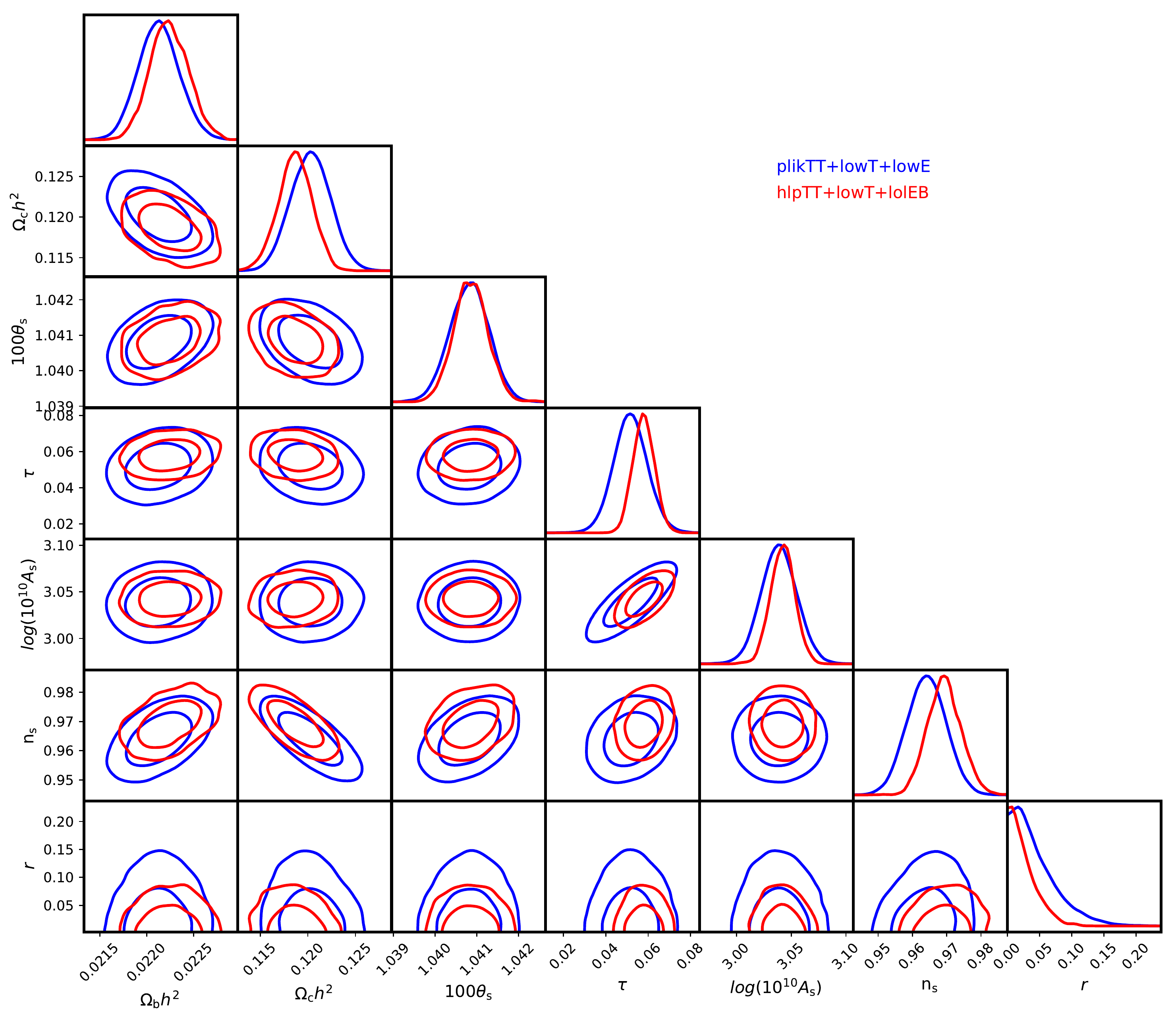} 
	\caption{Contour constraints for \LCDM+$r$ parameters for PR3 (\plik TT+\lowT+\lowE) and for PR4 (\hlp TT+\lowT+\lolEB).}
	\label{fig:triangle_lcdm}
\end{figure*}

Figure~\ref{fig:triangle_lcdm} shows the differences in parameter values obtained for PR3 (\plik TT+\lowT+\lowE, Planck Legacy Archive) and PR4 (\hlp TT+\lowT+\lolEB, this analysis) when both temperature and polarization data are included.  There are no significant differences. In general, uncertainties are slightly smaller with PR4.  

\section{Comparison with other \textit{BB} measurements}
\label{ann:BBplot}

\begin{figure*}[htbp!]
	\center
	\includegraphics[draft=\draft,width=0.9\textwidth]{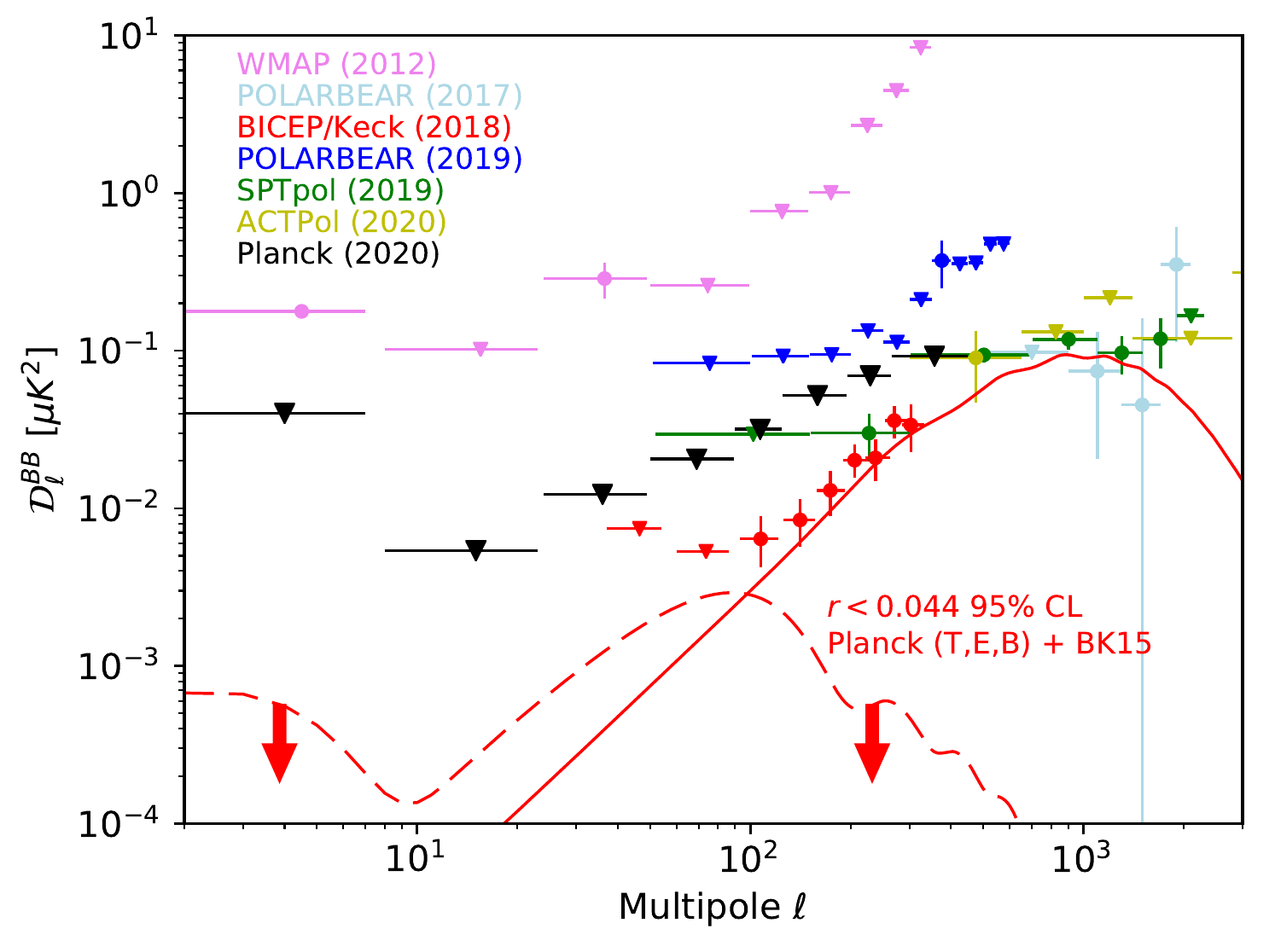} 
	\caption{$B$-mode polarization power spectrum measurements from \planck\ PR4 (this work), SPTpol \citep{Sayre20}, ACTPol \citep{Choi20}, POLARBEAR \citep{polarbear17,polarbear19}, the BICEP2/Keck Array \citep{Bicep2018limit}, and WMAP~\citep{bennett2012}. Uncertainties are 68\,\% confidence levels, while bandpowers compatible with zero at 95\% CL are plotted as upper limits. The solid red curve shows the $BB$ lensing spectrum based on the \planck\ 2018 \LCDM\ best-fit model \citep{planck2016-l06}. The dashed red curve corresponds to the $BB$ power spectrum for $r = 0.044$, the upper limit obtained in this work when combining \Planck\ ($TT$, $EE$, and $BB$) with BK15.}
	\label{fig:BBplot}
\end{figure*}

\Planck\ $TT$, $EE$, and $BB$ power spectra all contribute to the constraints on $r$ derived in this paper.  Most previous limits on $r$ have been determined only from $BB$ measurements.  For the convenience of readers who might be interested in how \Planck\ $BB$ data {\it alone\/} compare to those of other experiments, Fig.~\ref{fig:BBplot} shows the \planck\ $BB$ bandpowers as upper limits at 95\% CL, along with other measurements from the BICEP2/Keck Array \citep{Bicep2018limit}, SPTpol \citep{Sayre20}, ACTPol \citep{Choi20}, POLARBEAR \citep{polarbear17,polarbear19}, and WMAP \citep{bennett2012}.

\end{document}